\documentclass[a4paper,11pt]{article}
\pdfoutput=1 
\usepackage{amsmath, amssymb, amscd}
\usepackage[top=3.5cm, bottom=3.5cm, left=3cm, right=3cm]{geometry}
\usepackage{setspace}
\usepackage{framed}
\usepackage{color}
\usepackage{verbatim}
\usepackage{graphicx}
\usepackage{epstopdf}
\usepackage [english]{babel}
\usepackage [autostyle, english = american]{csquotes}

\usepackage{bm}
\usepackage{cite}

\usepackage{hyperref} 

\MakeOuterQuote{"} 

\numberwithin{figure}{section}
\numberwithin{equation}{section}

\newcommand{\be}{\begin{equation}}
\newcommand{\ee}{\end{equation}}
\newcommand{\bea}{\begin{eqnarray}}
\newcommand{\eea}{\end{eqnarray}}

\newcommand\ie{\textit{i.e.}\ }
\newcommand\eg{\textit{e.g.}\ }
\newcommand\cf{\textit{cf.}\ }

\newcommand{\aka}{{a.k.a.}\ }

\newcommand{\half}{\tfrac{1}{2}}

\newcommand{\ph}{\varphi}
\newcommand{\hG}[2]{\hat{\Gamma}_{#1}^{#2}}
\newcommand{\hS}[1]{\hat{S}^{#1}}
\newcommand{\hbS}[1]{\hat{\cal S}^{#1}}
\newcommand{\bS}{{\cal S}^\Lambda}
\newcommand{\C}{\tilde{C}^\Lambda}
\newcommand{\Cp}{\mathring{C}^{\mathring{\Lambda}}}

\newcommand{\Cpp}{{C}^{\mathring{\Lambda}}_k}

\newcommand{\Lp}{{\mathring{\Lambda}}}
\newcommand{\php}{\mathring{\ph}}

\begin{document}
\begin{titlepage}

\begin{center}
{\huge \bf Solutions to the Reconstruction Problem in Asymptotic Safety}
\end{center}
\vskip1cm


\begin{center}
{\bf Tim R. Morris and Zo\"e H. Slade}
\end{center}

\begin{center}
{\it STAG Research Centre \& School of Physics and Astronomy,\\  University of Southampton,
Highfield, Southampton, SO17 1BJ, U.K.}\\
\vspace*{0.3cm}
{\tt  T.R.Morris@soton.ac.uk,\\ Z.Slade@soton.ac.uk}
\end{center}

\abstract{Starting from a full renormalised trajectory for the effective average action (\aka infrared cutoff Legendre effective action) $\Gamma_k$, we explicitly reconstruct  corresponding bare actions, formulated in one of two ways. The first step is to construct the corresponding Wilsonian effective action $S^k$ through a tree-level expansion in terms of the vertices provided by $\Gamma_k$. It forms a perfect bare action giving the same renormalised trajectory. A bare action with some ultraviolet cutoff scale $\Lambda$ and infrared cutoff $k$ 
necessarily produces an effective average action $\Gamma^\Lambda_k$ that depends on both cutoffs, but if the already computed $S^\Lambda$ is used, we show how $\Gamma^\Lambda_k$ can also be computed from $\Gamma_k$ by a tree-level expansion, and that $\Gamma^\Lambda_k\to\Gamma_k$ as $\Lambda\to\infty$. Along the way we show that Legendre effective actions with different UV cutoff profiles, but which correspond to the same Wilsonian effective action, are related through tree-level expansions. All these expansions follow from Legendre transform relationships that can be derived from the original one between $\Gamma^\Lambda_k$ and $S^k$.}



\end{titlepage}
\tableofcontents

\section{Introduction}

One attempted route to a quantum theory of gravity is through the asymptotic safety programme \cite{Weinberg:1980,Reuter:1996}.  Although quantum gravity based on the Einstein-Hilbert action is plagued by ultraviolet infinities that are perturbatively non-renormalisable (implying the need for an infinite number of coupling constants), a sensible theory of quantum gravity might be recovered if there exists a suitable ultraviolet fixed point \cite{Weinberg:1980}. 

An appropriate technique to investigate this proposal is provided by the functional Renormalisation Group (RG) expressed in terms of the effective average action \cite{Wetterich:1992,Morris:1993}, a framework that has been applied in a wide variety of contexts, see \eg the reviews \cite{Morris:1998,Bagnuls:2000,Berges:2000,Gies:2006wv,Pawlowski:2005xe}. Beginning with ref. \cite{Reuter:1996}, there is a wealth of literature investigating asymptotic safety in this way. For  reviews and introductions see \cite{Reuter:2012,Percacci:2011fr,Niedermaier:2006wt,Nagy:2012ef,Litim:2011cp}, and for recent advances see for example 
\cite{Falls:2014tra,Dietz:2015owa,Demmel:2014hla,Dona:2014pla,Falls:2015cta,Saltas:2014cta,Eichhorn:2015bna,Becker:2014pea,Percacci:2015wwa,Ohta:2015efa}. In the vast majority of this work the RG flow equation takes the generic form \cite{Wetterich:1992}:
\begin{equation}
	\label{R-flow}
	\frac{\partial}{\partial k}\hat{\Gamma}_{k}[\varphi]=\frac{1}											{2}\text{tr}\bigg[\bigg(R_{k}+\frac{\delta^{2}\hat{\Gamma}_{k}}										{\delta\varphi\delta\varphi}\bigg)^{\!-1}\frac{\partial R_{k}}{\partial k}\bigg]\,.
	\end{equation}
In here tr is a (Euclidean) space-time trace, and for illustrative purposes we have written the flow equation for a single-component scalar field $\ph$. $\hat\Gamma_k$ is the effective average action, which is the Legendre effective action (\ie generator of one-particle irreducible diagrams) but where an infrared (IR) cutoff term $\half\int_p \ph(p)R_k(p^2)\ph(-p)$ has been added to suppress modes with momenta below the IR scale $k$. By varying 
$k$, a non-perturbative RG flow is generated with the property that for $k \rightarrow 0$ the information contained in the full functional integral is recovered. 

Of course to apply this technique to gravity, we need instead to work with some metric degrees of freedom. Typically this is
the gauge-fixed metric $g_{\mu\nu}$,  and thus necessarily also a background metric ${\bar g}_{\mu\nu}$ and ghost fields ${\cal C}_\mu$ and ${\bar{\cal C}}_\mu$. Actually, in this paper almost none of this extra structure plays a r\^ole. Therefore in the following discussions we will phrase all arguments in terms of this single-component scalar field $\ph$. It is straightforward to adapt the equations to more fields with indices and/or different statistics as required.  We make more comments on this in the conclusions.

As will be reviewed in sec. \ref{sec:SandGamma}, the flow equation \eqref{R-flow} is derived from the partition function and therefore strictly speaking should be subject to the same overall ultraviolet (UV) cutoff $\Lambda$ that is required to make sense of the functional integral \cite{Morris:1993}. Providing the IR cutoff profile $R_k(p^2)$ varies sufficiently fast, the flow \eqref{R-flow} itself however receives support only from finite $|p|/k$ and thus is well defined in the limit that the UV cutoff is removed ($\Lambda\to\infty$). In other words we can solve for the flow equations ``directly in the continuum'' (as already emphasised in ref. \cite{Morris:1993}). In fact in general this is crucial to its use since this allows everything to be expressed in terms of flowing dimensionless couplings $g_i(k)$ with respect to the \emph{single} dimensionful parameter $k$, \ie to recover the power of the Wilsonian RG \cite{Wilson:1973}. In particular only in this way can we find fixed points with respect to $k$ (implying the absence of any other dimensionful parameter), and construct the continuum limit in the standard way envisaged in asymptotic safety literature, namely {via}  the full \emph{renormalised trajectory}, starting from the UV fixed point $g_i=g^*_i$ at $k=\infty$ and flowing down to $k=0$ where the IR cutoff is finally removed. The renormalised trajectory is then parametrised by the running renormalised couplings $g_{i\in{\cal R}}(k)$, where ${\cal R}$ is the set of (marginally) relevant directions.\footnote{Throughout this Introduction we will rely on such standard Wilsonian concepts, also including Wilsonian effective actions, perfect actions, and tuning general bare actions to a continuum limit. For a review see \eg \cite{Morris:1998}.}

In this way, we dispense with the need to define a bare action $\hat{\mathcal{S}}^{ \Lambda}[\ph]$ at the overall cutoff scale $\Lambda$ and concomitant tuning required to reach the continuum limit.  However as emphasised by Manrique and Reuter \cite{Manrique:2008zw,Manrique:2009tj} this leaves us with a problem, dubbed by them ``the reconstruction problem'',\footnote{See also ref. \cite{Vacca:2011fx}.} since potentially we need access to some bare action to obtain the microscopic degrees of freedom, and from there study possible Hamiltonian formulations, understand more directly properties of the constructed quantum field theory such as constraints and local symmetries, make more direct contact with perturbative approaches, and finally more directly compare this to other approaches that are formulated at the microscopic level, such as canonical quantisation, loop quantum gravity or Monte Carlo simulations \cite{Ashtekar:2004eh,Thiemann:2007zz,Rovelli:2008zza,Ambjorn:2012jv,Ashtekar:2014kba,Hamber:2015jja,Liu:2015bwa}.

In order to make the issue more concrete, Manrique and Reuter consider the following situation \cite{Manrique:2008zw,Manrique:2009tj}. They regulate the functional integral by using a sharp cutoff $\Lambda$, such that the integration is restricted to only those modes propagating with momentum $|p|\equiv\sqrt{p^{2}}\leq \Lambda$, and consider either a generic IR cutoff profile $R_k$ or the optimised cutoff profile defined as $R_{k}(p)=(k^2-p^2)\theta(k^2-p^2)$\cite{opt1,Litim:2001,opt3}. Now there are actually two problems to confront. Firstly the resulting UV regulated flow equation is not that of \eqref{R-flow}: it depends on two cutoffs now, namely the IR cutoff $k$ and the UV cutoff $\Lambda$. Any solution $\hat{\Gamma}^\Lambda_k$ to this flow equation,\footnote{Notice that we denote an UV cutoff parameter with a superscript and an IR cutoff parameter with a subscript. We will use this pictorial guide throughout the paper.}   therefore also depends on these two cutoffs.\footnote{In ref. \cite{Manrique:2008zw} it is claimed that for the optimised cutoff this dependence disappears in the sense that providing we restrict flows to $k\le\Lambda$, we can consistently set $\hG{k}\Lambda[\ph] = \hG{k}{}[\ph]$. In fact this is not correct, as explained in appendix \ref{app:UV}.} This is a severe 
issue because, as explained above, it is crucial in practice that the flow equations are solved directly in the continuum where only one cutoff scale is operating. There is also a second  
severe issue. Even when $k=\Lambda$, there is still a functional integral to do,  albeit threshold-like, being only over modes with an effective mass of order the overall cutoff $\Lambda$. Thus the effective action $\hat{\Gamma}^\Lambda_{k=\Lambda}$ is related to a bare action $\hat{\mathcal{S}}^{ \Lambda}[\ph]$ in a way which cannot in practice be calculated exactly, and moreover we then need to invert this relation in order to find $\hat{\mathcal{S}}^{ \Lambda}[\ph]$ in terms of $\hat{\Gamma}^\Lambda_{k=\Lambda}$.  At the one-loop level, the partition function can be evaluated by steepest descents \cite{Manrique:2008zw}:\footnote{
All momenta  should be understood to be cutoff from above by $\Lambda$, including that in the momentum integral implied by the space-time trace. The mass parameter $M$ introduced in ref. \cite{Manrique:2008zw} will play no significant r\^ole here so will be neglected. Also in contrast to ref. \cite{Manrique:2008zw}, we will not make the momenta discrete by compactifying on a torus.}
\begin{equation}
	\label{MandR.soln}
	\hat{\Gamma}^\Lambda_{k=\Lambda}[\varphi]-\hat{\mathcal{S}}^{ \Lambda}[\varphi]=\frac{1}{2}\text{tr}\,\text{ln}			\Big\{\hat{\mathcal{S}}^{ \Lambda (2)}[\varphi]+R_{\Lambda}\Big\}\,,
	\end{equation}
where $\hat{\mathcal{S}}^{ \Lambda (2)} = \delta^2 \hat{\mathcal{S}}^{ \Lambda}/\delta\varphi\delta\varphi$ is the Hessian of the bare action. Unfortunately in the interesting case of asymptotic safety the theory is strongly interacting at these scales, with all couplings ${\cal O}(1)$ times the appropriate power of $\Lambda$, and thus one loop is not a good approximation. Furthermore even with this approximation it is not straightforward to invert the relation to find $\hat{\mathcal{S}}^{ \Lambda}[\varphi]$ in terms of $\hat{\Gamma}^\Lambda_{k=\Lambda}$. This then is the reconstruction problem.

Actually a practical prescription for reconstructing the bare action can be given, closely based on results from ref. \cite{Morris:1993}. (Aspects of reconstruction were already treated there at the end of sec. 3 and in the Conclusions.) As we will see, this prescription solves both of the above issues for a wide range of cutoffs by utilising a kind of duality relation between a Wilsonian effective action $\hat{S}^k$ and the effective average action $\hG{k}{}$. In particular it also provides a map between an effective multiplicative UV cutoff $C^k(p^2)$ and the IR cutoff $R_k(p^2)$. If an overall UV cutoff at $\Lambda$ is in place (of some form, not necessarily the sharp cutoff considered in ref. \cite{Manrique:2008zw}), then this is also involved in the map. Although such an overall UV cutoff necessarily modifies the flow equation \eqref{R-flow}, it is possible and natural to choose that the effective UV cutoff $C^k$ depends only on $k$. The central point is that since the Wilsonian effective action is already an action which is fully regularised in the UV by $C^k$, it can be used as a bare action. 

Since $\hS{k}$ depends on only one scale, namely $k$, it can also display all the required RG properties. In particular 
in the continuum limit the full trajectory $\hS{k}$ is then again the renormalised trajectory, but couched in this language, starting from the UV fixed point\footnote{Here we commit a slight abuse of notation. Strictly in order for the action to reach a fixed point, we should change to the appropriate dimensionless variables. By $\hS*$ we actually mean the action such that it takes the fixed point form after such a transformation.} $\hS*$ in the far UV ($k\to\infty$) and extending down to $k\to0$. It follows that such an $\hS{k}$ is a continuum version of the \emph{perfect bare actions} explored \eg in ref. \cite{Hasenfratz:1993sp} since, as we review in sec. \ref{sec:SandGamma}, setting $\hbS\Lambda=\hS{k=\Lambda}$ to be the bare action (together with UV cutoff $C^{k=\Lambda}$) results in a partition function that is actually \emph{independent} of $\Lambda$ and thus in particular equal to the partition function obtained in the continuum limit $\Lambda\to\infty$.  

Unlike the map described in \eqref{MandR.soln}, the map between $\hS{k}$ and $\hG{k}{}$ is exact. Unlike the map \eqref{MandR.soln}, it is straightforward to explicitly construct it in either direction, via a tree-diagram expansion which can be developed vertex by vertex, as we will see in sec. \ref{sec:vertices}. It is also possible to solve the relation explicitly in approximations that go beyond an expansion in vertices. For example the duality relation remains exact in the Local Potential Approximation and thus at this level can be analysed exactly, both analytically and numerically \cite{Morris:2005ck,Bonanno1}.

It should thus be clear that constructing $\hbS{\Lambda}=\hS\Lambda$ in this way, already provides a practical solution to the reconstruction problem, since it furnishes a bare action that expresses the same asymptotically safe renormalised trajectory as $\hG{k}{}$.  

This still leaves a puzzle however, since it is not immediately clear how this solution should be related to the one-loop expression \eqref{MandR.soln}. Actually, as already emphasised, and proved in appendix \ref{app:UV}, a partition function regularised by some finite UV cutoff $\Lambda$ \emph{cannot} through the standard Legendre transform relations yield the continuum Legendre effective action $\hG{k}{}$. Instead it must give an effective average action $\hG{k}\Lambda$ that now also depends explicitly on $\Lambda$. Therefore $\hG{k}{}$ does not result from computing the partition function defined by using $\hbS{\Lambda}=\hS\Lambda$ together with the infrared cutoff $R_k$. 
If we want $\hbS\Lambda$ to be associated to $\hG{k}{}$ in this sense, then the best we can hope to achieve is to find a map from the continuum $\hG{k}{}$ to a pair $\{\hbS\Lambda, \hG{k}\Lambda\}$ consisting of a bare action 
and the resulting effective average action, such that $\hG{k}\Lambda\to\hG{k}{}$ as $\Lambda\to\infty$. In this paper we set out exactly such a map, again explicitly constructable vertex by vertex, and show how it is consistent with the one-loop formula \eqref{MandR.soln}.

This alternative prescription for reconstruction is set out precisely and in more detail in secs. \ref{sec:details} and \ref{sec:solving}. Here we briefly sketch the main steps. Assume we have found the appropriate renormalised trajectory $\hG{k}{}\equiv\hG{k}\infty$ of \eqref{R-flow}, where we emphasise that this solution corresponds to the case where the overall UV cutoff has been removed. Using the duality relation we construct the corresponding Wilson effective action $\hS{k}$ together with its associated effective UV cutoff $C^k$. We set this to be a bare action at $k=\Lambda$, \ie $\hat{\mathcal{S}}^\Lambda = \hS{k=\Lambda}$.  
However we replace the multiplicative cutoff $C^\Lambda$ with  $C^\Lambda_k = C^\Lambda - C^k$. 
As we will see, this cutoff has the property that it regularises both in the IR and the UV. It also has the properties that it is $C^\Lambda$ in the limit $k\to0$, and provides exactly $R_k$ in the limit $\Lambda\to\infty$.   In the standard way the partition function now yields an effective average action $\hG{k}\Lambda$. 
$\hG{k}\Lambda$ satisfies a UV regularised version of the flow equation \eqref{R-flow} with the property that as $\Lambda\to\infty$ it goes back to the original flow equation \eqref{R-flow}. However we do not need to solve this new flow equation, or do the functional integral, in order to construct $\hG{k}\Lambda$. It turns out that essentially the same duality relation allows us to construct  $\hG{k}\Lambda$ exactly from $\hG{k}{}$, again vertex by vertex 
or by other methods, as before \cite{Morris:1993}. 

Thus we have constructed an exact, explicit and calculable map from any continuum solution $\hG{k}{}\equiv\hG{k}\infty$ with its associated IR cutoff $R_k$, to the pair, $\hat{\mathcal{S}}^\Lambda$  and $\hG{k}\Lambda$, related in the standard way through a functional integral regularised in the UV and IR by $C^\Lambda_k$.
As advertised, this pair has the property that as $\Lambda\to\infty$, the regularised solution $\hG{k}\Lambda\to\hG{k}{}$. Since, given $\hG{k}{}$, everything is explicitly calculable, we see that this provides an alternative solution to the reconstruction problem.

The relation between the bare action $\hat{\mathcal{S}}^\Lambda$ and the `initial' UV value $\hG{k=\Lambda}\Lambda$ of this regularised effective average action, is particularly simple. They are simply equal. Together with the associated Wilsonian effective action we thus have the triple equality: $\hG\Lambda\Lambda = \hat{\mathcal{S}}^\Lambda = \hS\Lambda$, which moreover is dual to the original continuum solution $\hG{k}{}$ evaluated at $k=\Lambda$.

In sec. \ref{sec:solving} we show how this solution is consistent with the one-loop formula \eqref{MandR.soln}.  On the one hand by construction the multiplicative cutoff $C^\Lambda_{k}$ vanishes at $k=\Lambda$, which means effectively that the modified $R_k$ diverges at $k=\Lambda$. In this case we say that the UV and IR cutoffs are \emph{compatible}. As a consequence, apart from a field independent piece, \eqref{MandR.soln} implies that $\hG\Lambda\Lambda = \hat{\mathcal{S}}^\Lambda$, recovering our result. On the other hand if the UV and IR cutoffs are not compatible, there is still a functional integral to do at $k=\Lambda$. Then the formula \eqref{MandR.soln} supplies the approximate relation, valid to one loop. As we review in secs. \ref{sec:SandGamma} and \ref{sec:solving}, the Wilsonian effective action $\hS{k}$ can also be derived from the bare action $\hat{\mathcal{S}}^\Lambda$ via  a functional integral. In sec. \ref{sec:solving}, we show directly by the method of steepest descents that in the non-compatible case this functional integral yields at one loop an $\hS\Lambda$ which is precisely the one which is dual to the effective action given by the formula \eqref{MandR.soln}, proving consistency also in the non-compatible case. 

We have thus provided two solutions to the reconstruction problem. In the Conclusions, we emphasise that there are in fact infinitely many solutions, and sketch how some of these can be constructed starting with the explicit solutions given here.

The structure of the paper is then as follows. In the next section we give the definitions in order to set out precisely our two prescriptions for reconstructing a bare action. For the second prescription we use a special case of a remarkable relation proved in sec. \ref{sec:duality-Gamma}. There we prove another Legendre transform (\aka duality) relation between two effective average actions, or simply two Legendre effective actions, with different overall UV cutoff profiles but the same associated Wilsonian effective action. In sec. \ref{sec:SandGamma} we derive the main Legendre transform relation between Wilsonian effective actions and effective average actions, and show how these are in turn derived from the partition function, extending the results of ref. \cite{Morris:1993} to more general cutoff profiles. In sec. \ref{sec:vertices} we compute the vertices of the Wilsonian effective action $\hat{S}^k$ from $\hat{\Gamma}_k$ through the tree-level expansion implied by the duality relation. This expansion can also be used in the other direction and for the other duality relations simply by renaming propagators and vertices. In sec. \ref{sec:solving} we provide more detail on our second solution to the reconstruction problem and show how it is related to \eqref{MandR.soln}. In sec. \ref{sec:compatible} we give some examples of compatible cutoff profiles, and finally in sec. \ref{sec:conclusions} we summarise and draw our conclusions.

\section{Detailed prescription for reconstruction}
\label{sec:details}


Here we set out in detail the definitions of the quantities we need in order to set out precisely our two prescriptions for reconstructing a bare action, as sketched in the Introduction.  We will use a dot notation to denote integration over position or momentum space: 
\be
\label{dot1}
J\cdot\phi\equiv J_{x}\phi_{x} \equiv \int d^{d}x J(x)\phi(x)=\int \frac{d^{d}p}{(2\pi)^{d}} J(p)\phi(-p)\,.
\ee 
For bilinear terms we regard the kernel as a matrix, thus the following forms are equivalent:
\be
\label{dot2}
\phi\cdot\Delta^{\!-1}\!\!\cdot\phi \equiv \phi_{x}\Delta^{\!-1}_{xy}\phi_{y} \equiv\int \!\!d^{d}xd^{d}y\,\phi(x)\Delta^{\!-1}(x,y)\phi(y)=\int \frac{d^{d}p}{(2\pi)^{d}}\phi(p)\Delta^{\!-1}(p^{2})\phi(-p)\,.
\ee 
Note that when transforming to momentum space, Green's functions $G(p_{1},\cdots,p_{n})$ come with momentum conserving delta functions such that they are only defined for $p_{1}+\cdots+p_{n}=0$. Thus two-point functions are functions of just a single momentum $p=p_{1}=-p_{2}$.


Let us choose to define the interaction part of the effective average action to be the part obtained by splitting off a normalised massless kinetic term:
\be
\label{interactions}
\hat{\Gamma}_{k}[\ph]= \frac{1}{2}\varphi\cdot p^{2}\cdot \varphi+\Gamma_{k}[\ph]\,.
\ee
Note that for the purposes of this accounting we regard a mass term $\frac{1}{2}m^{2}\ph^{2}$ as contained within the interactions. The total effective action contains also the additive infrared cutoff:
\bea 
\label{LEAA-con}
\Gamma^{\text{tot}}_{k}[\ph] &=& \hat{\Gamma}_{k}[\ph]+\frac{1}{2}\varphi\cdot R_{k}\cdot \varphi\,, \\
\label{LEAA-alt}
&=& \Gamma_{k}[\ph]+
	\frac{1}{2}\varphi\cdot \left(\Delta_{k}\right)^{\!-1}\!\!\cdot \varphi\,,
\eea
where in the second line we have combined the massless kinetic term with the additive cutoff to form a  propagator with a multiplicative cutoff:
\be
\label{DeltaIR}
\Delta_k = \frac{C_k(p)}{p^2}\,,
\ee
such that
\be 
\label{IR-only}
C_k(p) = \frac{p^2}{p^2+R_k(p)}\,.
\ee
This provides the translation between multiplicative IR cutoff profiles and additive IR cutoff profiles, but explicitly uses the fact that the UV cutoff has been removed. Note that $C_k$ inherits from $R_k$ the properties that for $|p|<k$ it suppresses modes, and in particular $C_{k}(p)\to 0$ as $|p|/k\to0$, while for $|p|>k$, $C_k(p)\approx1$ and mostly leaves the modes unaffected and in particular  $C_{k}(p) \rightarrow1$ as $|p|/k\to\infty$.

As already mentioned in the Introduction, Manrique and Reuter choose to regularise the functional integral in the UV with an overall sharp cutoff such that all momenta $|p|\le\Lambda$ \cite{Manrique:2008zw}. This is equivalent to ensuring that the internal momentum running through any propagator is cut off so that this propagator vanishes for $|p|>\Lambda$. Both the ultraviolet regularisation and the infrared regularisation can therefore be carried by a multiplicative cutoff 
\be
\label{UVIRMR}
C^\Lambda_k(p) = \frac{p^2\theta(\Lambda-|p|)}{p^2+R_k(p)}\,,
\ee
which appears in the resulting effective action like so:
\be
\label{total-Gamma}
\Gamma^{\text{tot},\, \Lambda}_{k}[\ph]= \Gamma^\Lambda_{k}[\ph]+
	\frac{1}{2}\varphi\cdot \left(\Delta_{k}^\Lambda\right)^{\!-1}\!\!\cdot \varphi\,,
\ee
where 
\be
\label{DeltaUVIR}
\Delta_k^\Lambda = \frac{C^\Lambda_k(p)}{p^2}\,.
\ee
We have noted that the effective average action now depends also on the overall UV cutoff $\Lambda$. We recover the previous case when the UV cutoff is removed: $C_k(p)\equiv C^\infty_k(p)$. 

As already emphasised in the Introduction, our constructions go through for much more general UV cutoffs, providing that the UV and IR cutoffs are always implemented together, multiplicatively, as defined via the above relations \eqref{total-Gamma} and \eqref{DeltaUVIR}.
As we recall in sec. \ref{sec:SandGamma}, the flow equation for the interactions then takes the general form
\begin{equation}
	\label{Gamma.flow}
	\frac{\partial}{\partial k}\Gamma^\Lambda_{k}[\varphi]=-\frac{1}{2}\text{tr}\bigg[\bigg(1+\Delta^\Lambda_k\cdot 						\frac{\delta^{2}\Gamma^\Lambda_{k}}{\delta\varphi\delta\varphi}\bigg)^{\!-1}\frac{1}									{\Delta^\Lambda_k}\frac{\partial\Delta^\Lambda_k}{\partial k}\bigg]\,.
	\end{equation}
By recasting the right hand side in terms of $\left(\Delta^\Lambda_k\right)^{\!-1}$, and using $1/$\eqref{DeltaIR}, $1/$\eqref{IR-only}, and \eqref{interactions}, it is easy to see that in the limit $\Lambda\to\infty$ this flow equation gives back \eqref{R-flow}.

Now we define in precisely the same way both the bare interactions $\mathcal{S}^\Lambda[\phi]$ and Wilsonian interactions $S^k[\Phi]$:
\be
\label{other-interactions}
\hbS\Lambda[\phi]= \frac{1}{2}\phi\cdot p^{2}\cdot \phi+\mathcal{S}^\Lambda[\phi]\,,\qquad 
\hat{S}^{k}[\Phi]= \frac{1}{2}\Phi\cdot p^{2}\cdot \Phi+S^{k}[\Phi]\,.
\ee
(We choose different symbols for the fields in each case, for convenience as will become clear later.) We define the total bare action to include also the UV cutoff profile and thus
\be 
\label{total-bare}
\mathcal{S}^{\mathrm{tot},\Lambda}[\phi] = \mathcal{S}^\Lambda[\phi]+\frac{1}{2}\phi\cdot \left(\tilde{\Delta}^{\Lambda}\right)^{-1}\!\!\cdot \phi\,,
\ee
where 
\be 
\label{DeltaTotalUV}
\tilde{\Delta}^\Lambda = \frac{\C(p)}{p^2}\,.
\ee
For the sharp cutoff case 
\be
\label{sharpUV}
\C(p)=\theta(\Lambda-|p|)\,,
\ee 
but again we emphasise that the UV cutoff profile can be more general and we will in general take it to be so. All  we then require is that for $|p|<\Lambda$, $\C(p)\approx1$ and mostly leaves the modes unaffected and in particular $\C(p) \to 1$ for $|p|/\Lambda\to0$, while for $|p|>\Lambda$ it suppresses modes, and in particular for $|p|/\Lambda\to\infty$, $\C(p) \to0$  sufficiently fast to ensure that all momentum integrals are regulated in the ultraviolet.
Finally, the total Wilsonian effective action can be written
\be 
\label{total-Wilsonian}
S^{\mathrm{tot},k}[\Phi] = S^k[\Phi] + \frac{1}{2}\Phi\cdot (\Delta^{k})^{\!-1}\!\!\cdot \Phi\,,
\ee
where 
\be 
\label{DeltaUV}
\Delta^k = \frac{C^k(p)}{p^2}\,,
\ee
and $C^k(p)$ is an ultraviolet cutoff profile for this effective action and effective partition function, which regularises at scale $k$. $C^k(p)$ has to satisfy the same conditions as $\C(p)$ above (with the replacement $\Lambda\mapsto k$ of course).
Since the functional integral with this action $S^{\mathrm{tot},k}$ is therefore already completely regularised in the ultraviolet, there is no need for any dependence on the overall UV cutoff $\Lambda$. We will therefore choose $C^k(p)$ to depend only on the one cutoff scale $k$ as already indicated, and indeed apart from obeying the same general conditions,  the profiles $C^k$ and $\C$ will otherwise be unrelated. However we will require one `sum rule' relation between these three profiles:\footnote{This goes beyond the sum rule introduced in ref. \cite{Morris:1993} since we now allow $\C$ to be unrelated to $C^k$.}
\be 
\label{sum rule}
C^\Lambda_k(p)+C^k(p) = \tilde{C}^\Lambda(p)\,.
\ee
For example, from \eqref{UVIRMR}, \eqref{sharpUV}, and \eqref{sum rule}, we can deduce the  UV cutoff profile for the Wilsonian effective action which is implied by the regularisation used in ref. \cite{Manrique:2008zw}:
\be
\label{MR-Wilson}
C^k(p) = \left(1-\frac{p^2}{k^2}\right)\theta(k-|p|)
\ee
(where $k<\Lambda$).
We see that it behaves sensibly as a UV cutoff profile and actually depends only on the one cutoff scale as required. 

Thus \eqref{UVIRMR}, \eqref{sharpUV} and \eqref{MR-Wilson} provide an example of a consistent set of cutoffs satisfying the sum rule \eqref{sum rule}.
However as noted in the Introduction, they are not compatible, in the sense that when the IR and UV cutoffs meet, $C^\Lambda_\Lambda$ does not vanish.
Examples of cutoff profiles satisfying \eqref{sum rule} that do also satisfy this compatibility condition are given in sec. \ref{sec:compatible}.



In general we can use \eqref{sum rule} to \emph{define} $C^\Lambda_k(p)=\C(p)-C^k(p)$. Since  $\Lambda>k$, the general properties given above for $C^k$ and $\C$ ensure that it behaves as a multiplicative UV cutoff at $\Lambda$ and multiplicative IR cutoff at $k$, as  required. Thus for $|p|>\Lambda$ modes are suppressed such that as $|p|/\Lambda\to\infty$, $C^\Lambda_k(p)\to0$ sufficiently fast that all momentum integrals are UV regulated. For $k<|p|<\Lambda$, $C^k(p)$ is small (vanishingly so for $|p|\gg k$) while $\C(p)\approx1$, and thus $C^\Lambda_k(p)\approx1$ and mostly leaves modes unaffected. For $k\ll |p|\ll \Lambda$, $C^\Lambda_k(p)$ will be very close to one. Finally for $|p|<k$, $C^k(p)\approx1$ and $\C(p)$ is close to one (very close for $k\ll\Lambda$) and thus $C^\Lambda_k(p)\approx0$ suppresses modes, while for $|p|/k\to0$, since both $C^k(p)\to1$ and $\C(p)\to1$, we have that $C^\Lambda_k(p)\to0$ thus providing the expected IR cutoff $k$.

By adding the infrared cutoff profile to the bare action in order to generate the effective average action in the usual way, we equivalently change the multiplicative cutoff profile $\C$ into one that depends on both $\Lambda$ and $k$. We have already anticipated in our discussion of $\Gamma^\Lambda_k$ that the new multiplicative cutoff profile is $C^\Lambda_k$. Thus the bare action becomes
\be 
\label{total-bare-k}
\mathcal{S}^{\mathrm{tot},\Lambda}_k[\phi] = \mathcal{S}^\Lambda[\phi]+\frac{1}{2}\phi\cdot \left({\Delta}^{\Lambda}_k\right)^{\!-1}\!\!\cdot \phi\,.
\ee
It is this bare action that generates \eqref{total-Gamma} in the usual way and leads to the UV modified flow equation \eqref{Gamma.flow}. (Note that the total bare action then necessarily depends on both cutoffs. The bare interactions $\bS$ however do not, and indeed consistent with the usual philosophy of renormalisation they should be taken to depend only on the UV modification.)

Now we can state the duality in its general form:
\begin{equation}
	\label{duality-gen}
	S^{k}[\Phi]=\Gamma_{k}^\Lambda[\varphi]+\frac{1}{2}(\varphi-\Phi)\cdot \left(\Delta_{k}^{\Lambda}\right)^{\!-1}\!\!\cdot (\varphi-\Phi)\,.
\end{equation}
This is a Legendre transform relation that maps between two apparently very different pictures of the exact RG \cite{Morris:1993}. On the one hand we have the effective average action which flows with respect to an IR cutoff $k$ as in \eqref{Gamma.flow} (or in the limit $\Lambda\to\infty$, as in  \eqref{R-flow}) and on the other hand we have a Wilsonian effective action whose interactions flow with respect to an effective UV cutoff $k$:
	\begin{equation}
	\label{S.flow}
	\frac{\partial}{\partial k}S^k[\Phi]=\frac{1}{2}\frac{\delta S^k}{\delta\Phi}\cdot \frac{\partial\Delta^k}{\partial k}\cdot			\frac{\delta S^k}{\delta\Phi}-\frac{1}{2}\text{tr}\bigg(\frac{\partial\Delta^k}{\partial k}\cdot \frac{\delta^{2}S^k}			{\delta\Phi\delta\Phi}\bigg)\,,
	\end{equation}
this being the Polchinski flow equation \cite{Polchinski:1983gv}, which can also be regarded as equivalent to Wilson's original equation. See sec. \ref{sec:SandGamma} and refs. \cite{Morris:1993,Morris:1998}.\footnote{For a more careful comparison between Wilson's and Polchinski's versions see ref. \cite{Bervillier:2013kda}.} As we also review in sec. \ref{sec:SandGamma}, and outlined in the Introduction, the original partition function with bare action \eqref{total-bare} can be exactly re-expressed as a partition function with the bare action replaced with the Wilsonian one \eqref{total-Wilsonian}, which is thus a so-called ``perfect action''. 

In particular if we have an effective average action solution $\Gamma_k$  to the continuum flow equation \eqref{R-flow} such that it exists for all $0< k<\infty$, we can construct $S^k$ by using \eqref{duality-gen} with the identifications $\Gamma_k\equiv\Gamma^\infty_k$, and $\Delta_k \equiv \Delta^\infty_k$ as in \eqref{DeltaIR} and \eqref{IR-only}:
\begin{equation}
	\label{duality-con}
	S^{k}[\Phi]=\Gamma_{k}[\varphi]+\frac{1}{2}(\varphi-\Phi)\cdot \left(\Delta_{k}\right)^{\!-1}\!\!\cdot (\varphi-\Phi)\,.
\end{equation}
$S^k$ can then be constructed from this for example vertex by vertex as in sec. \ref{sec:vertices}.  

We can then reconstruct the partition function $Z$ even in this continuum limit, 
by using the ``perfect'' bare action \eqref{total-Wilsonian} with $k$ set to some `initial' upper scale of our choice, $k=\Lambda$ for example. Note that as required such an action has the same structure as the general form of the bare action \eqref{total-bare}, and indeed just involves the replacements $\C\mapsto C^\Lambda$, and $\bS\mapsto S^\Lambda$. The new UV cutoff profile $C^k(p)=1-C_k(p)$ as follows from \eqref{sum rule} with $\tilde{C}^\infty \mapsto 1$. This then provides our first solution to the reconstruction problem.

Note that such a bare action, and thus partition function, does not incorporate an infrared cutoff $R_k$ and thus there is no connection to the effective average action $\Gamma_k$ through the standard route of taking a Legendre transform of $\ln Z$. If we add the infrared cutoff term to this bare action, we still do not recover $\Gamma_k$ this way.  As emphasised in the Introduction and appendix \ref{app:UV}, it is impossible to recover the continuum effective average action this way since the result is a $\Gamma^\Lambda_k$ that necessarily now depends on both cutoffs. It is possible however to construct a map from the continuum solution $\Gamma_k$ to a pair $\{\bS, \Gamma_{k}^\Lambda\}$, where $\Gamma^\Lambda_k$ is related to $\bS$ in the usual way, and such that as $\Lambda\to\infty$ we have $\Gamma_{k}^\Lambda\to\Gamma_k$. This is our second solution to the reconstruction problem. 

To construct this solution we specialise to cutoffs that are \emph{compatible}, as defined in the Introduction. This means that the overall UV cutoff $\C = C^{k=\Lambda}$ is identical to the effective UV cutoff set at scale $k=\Lambda$. We again take the bare interactions to be the Wilsonian interactions $\bS=S^{k=\Lambda}$ computed as above. The total bare action $\mathcal{S}^{\mathrm{tot},\Lambda}_k$, now with the infrared cutoff in place, is given as it should be, by \eqref{total-bare-k}, \ie regularised by $C^\Lambda_k$. The corresponding partition function yields by the standard construction a $\Gamma^\Lambda_k$ which satisfies the UV modified flow equation \eqref{Gamma.flow}. 

This provides the map we required. Note that $C^\Lambda_\Lambda(p)$ vanishes for all $p$, by the sum rule formula \eqref{sum rule}. Thus \eqref{duality-gen} implies that 
\be
\label{initial}
\Gamma^\Lambda_\Lambda[\ph]=S^\Lambda[\ph] 
\ee
(and $\ph=\Phi$) as can be seen either directly from the fact that $1/\Delta_{\Lambda}^{\Lambda}(p)$ is infinite for all $p$, or more carefully by first solving the Legendre transform relation as done for the continuum version in \eqref{soln2}. The UV boundary condition \eqref{initial} for the flow \eqref{Gamma.flow} is therefore particularly simple, and is a triple equality since the right hand side is also the bare interactions. More details are given in sec. \ref{sec:solving}.

We do not need to compute the functional integral or solve the flow \eqref{Gamma.flow} to find $\Gamma_k^\Lambda$ however. This can also be constructed vertex by vertex from the original continuum $\Gamma_k$ using the same recipe as in sec. \ref{sec:vertices}. The clue is hidden in a remarkable property of the duality relation \eqref{duality-gen}. Note that by construction $S^k$ need have no dependence on $\Lambda$. (It is just a solution to \eqref{S.flow} which also has no dependence on $\Lambda$.) Therefore if we choose to keep $S^k$ fixed, the duality relation \eqref{duality-gen} actually implies that the right hand side is independent of the choice of overall UV cutoff $\C$, and in particular that it is independent of $\Lambda$. As we show in the next section, this implies that the two $\Gamma$s are related by
\be
\label{duality-Gamma-con}
\Gamma^\Lambda_k[\Phi] = \Gamma_k[\ph]+\frac{1}{2}(\ph-\Phi)\cdot \left(\Delta_{\Lambda}\right)^{\!-1}\!\!\cdot (\ph-\Phi)\,,
\ee
where the notation for the inverse propagator on the right hand side indicates that it is regularised in the infrared by $C_\Lambda:=C_{k=\Lambda}$. Comparing \eqref{duality-Gamma-con} to \eqref{duality-con}, we see that the vertices of $\Gamma^\Lambda_k[\Phi]$ are thus given by those of $S^k$ in the recipe set out in sec. \ref{sec:vertices},
providing we make the replacement $\Delta_k\mapsto\Delta_\Lambda$. Of course it then follows that the same tree-diagram expansion illustrated in fig. \ref{fig:vertices} is  also correct for $\Gamma^\Lambda_k[\Phi]$ after this replacement.

\section{Proof of a duality relation between effective actions with different UV regularisations.}
\label{sec:duality-Gamma}
We will consider a more general case and then specialise to \eqref{duality-Gamma-con}, since the proof is just as simple. We thus go back to the UV regularised form \eqref{duality-gen} of the duality relation between the Wilsonian interactions $S^k$ and the effective average action $\Gamma^\Lambda_k$. Now consider an alternative overall UV cutoff $\Cp(p)$ in place of $\C(p)$, where for generality we change both the profile form $\mathring{C}$ and the magnitude $\Lp$. Without loss of generality we can assume $\Lp>\Lambda$ however. We choose to keep the same effective UV cutoff $C^k$ and therefore through the sum rule relation \eqref{sum rule} we define an alternative joint regulator profile $\Cpp = \Cp - C^k$. Again, providing $\Cp$ is chosen to behave sensibly as a UV cutoff, as discussed below \eqref{sharpUV}, $\Cpp$ will behave correctly in the UV and infrared, as discussed below \eqref{MR-Wilson}.
Relabelling \eqref{duality-gen} in the obvious way, we evidently therefore have the alternative duality relation:
\begin{equation}
	\label{duality-gen-2}
	S^{k}[\Phi]=\Gamma_{k}^\Lp[\php]+\frac{1}{2}(\php-\Phi)\cdot \left(\Delta_{k}^{\Lp}\right)^{\!-1}\!\!\cdot (\php-\Phi)\,.
\end{equation}
As observed in the previous section, $S^k$ is not forced to have any dependence on these overall cutoffs. Since $S^k$ satisfies a flow equation \eqref{S.flow} which itself is independent of these cutoffs we can choose to keep the same solution $S^k$ after these changes.
Eliminating the left hand side we thus have the relation
\be
\label{eliminated-S}
\Gamma_{k}^\Lambda[\varphi]+\frac{1}{2}(\varphi-\Phi)\cdot \left(\Delta_{k}^{\Lambda}\right)^{\!-1}\!\!\cdot (\varphi-\Phi)
=\Gamma_{k}^\Lp[\php]+\frac{1}{2}(\php-\Phi)\cdot \left(\Delta_{k}^{\Lp}\right)^{\!-1}\!\!\cdot (\php-\Phi)\,.
\ee
This is a Legendre transform relation in which all three fields can be varied independently. Varying $\Phi$ we thus have
\be 
\label{Phi-elimination}
\left[\left(\Delta_{k}^{\Lp}\right)^{\!-1}-\left(\Delta_{k}^{\Lambda}\right)^{\!-1}\right]\Phi = \left(\Delta_{k}^{\Lp}\right)^{\!-1}\php-
\left(\Delta_{k}^{\Lambda}\right)^{\!-1}\ph\,.
\ee
Define $C^\Lp_\Lambda = \Cp-\C = \Cpp-C^\Lambda_k$, where the second equality follows from the sum rule \eqref{sum rule}. Given the general behaviour of its component parts, $C^\Lp_\Lambda$ is a multiplicative cutoff profile that is cutoff in the UV by $\Lp$ and in the IR by $\Lambda$, with properties discussed below \eqref{MR-Wilson}. Thus also define $\Delta^\Lp_\Lambda(p)=C^\Lp_\Lambda(p)/p^2$. Then \eqref{Phi-elimination} can be rearranged to give
\bea
\ph-\Phi &=& \frac{\Delta^\Lambda_k}{\Delta^\Lp_\Lambda}\cdot(\php-\ph)\,, \\
\php-\Phi &=& \frac{\Delta^\Lp_k}{\Delta^\Lp_\Lambda}\cdot(\php-\ph)\,.
\eea
Substituting these back into \eqref{eliminated-S} gives us the desired general duality relation between effective average actions with different UV cutoffs:
\be 
\label{duality-Gamma-gen}
\Gamma^\Lambda_k[\ph] = \Gamma^\Lp_k[\php] + \frac{1}{2}(\ph-\php)\cdot \left(\Delta_\Lambda^{\Lp}\right)^{\!-1}\!\!\cdot (\ph-\php)\,.
\ee
An alternative proof of this relation is given in ref. \cite{Morris:1993}, by demonstrating directly that this transformation turns the flow equation \eqref{Gamma.flow} into the equivalent one for $\Gamma^\Lp_k$. Ref. \cite{Morris:1993} however specialised to the case where only the scale $\Lambda\mapsto\Lp$ changes. As we see here the relation is more general including also the option to change the form of the cutoff profile. 

It is remarkable that such a generalised Legendre transformation relationship exists between two effective average actions regularised in the UV with different cutoff profiles, $\C$ versus $\Cp$. To drive the point home, note that we can take the limit $k\to0$ and then this is a Legendre transform relation between two standard Legendre effective actions regularised in different ways in UV of our choosing. This latter result is therefore significant in general, not just within the context of functional RG. As we see explicitly in sec. \ref{sec:vertices}, it implies that the vertices of two effective actions are related by tree diagram expansions which can be constructed exactly. 

Since a change in regularisation obviously affects the loop integrals in the quantum corrections, this result looks surprising at first sight. However note that the key to the relation is that the Wilsonian effective action \eqref{total-Wilsonian} is unchanged. Since $S^{{\rm tot},k}$ is ultimately derived from a functional integral that depends on the bare action \eqref{total-bare-k} 
(see \eqref{partition to S}),
which most certainly does depend on the form of the overall UV cutoff, the change from $\C$ to $\Cp$ implies a change of bare interactions $\bS\mapsto \mathring{{\cal S}}^\Lp$ sufficient to completely compensate for this when computing $S^{{\rm tot},k}$. We make further comments on this map in the conclusions. Although it makes no change to the Wilsonian effective action computed with these methods it leaves a remnant change to the Legendre effective action (with or without an IR cutoff $k$) which is summarised in the duality relation \eqref{duality-Gamma-gen}. 

In the special case where $\C = C^{k=\Lambda}$ and $\Cp = C^{k=\Lp}$, \ie where the UV scale changes but not the form of the cutoff, which is furthermore fixed to be the Wilsonian one, we have the situation already analysed in ref. \cite{Morris:1993}. Then the bare interactions change only trivially in that in each case ($k=\Lambda,\Lp$) the bare interactions are just equal to the Wilsonian interactions at that scale ${\cal S}^k = S^k$ as determined through the flow equation \eqref{S.flow}.

Finally, let us choose $\C=C^{k=\Lambda}$ and send $\Lp\to\infty$. Then $\Gamma^\Lp_k\to\Gamma_k$ and $C_\Lambda^{\Lp}\to1-C^\Lambda= C_\Lambda$, where we have used $\Cp\to1$ and \eqref{sum rule}. Thus with these changes, \eqref{duality-Gamma-gen} becomes the equation \eqref{duality-Gamma-con} we set out to prove.



\section{The Wilsonian Effective Action {\it versus} the Legendre Effective Action}
\label{sec:SandGamma}

In this section we recall most of the steps that give rise to the exact relationship \eqref{duality-gen} between the Wilsonian effective action and the Legendre effective action. They are adapted here from ref. \cite{Morris:1993} both because the relationship goes marginally beyond what was proven there and also because they underpin the claims in the rest of the paper.

We consider the functional integral for a scalar field $\phi(x)$ in a $d$-dimensional Euclidean spacetime:
	\begin{equation}
	\label{func.int}
	Z^\Lambda[J]=\int\!\!\mathcal{D}\phi\,\mathrm{e}^{
	-\mathcal{S}^{\mathrm{tot},\Lambda}[\phi] +J\cdot\phi}
	=\int\!\!\mathcal{D}\phi\,\mathrm{e}^{-\frac{1}{2}\phi\cdot\left(\tilde{\Delta}^\Lambda\right)^{\!-1}\!\!\cdot\phi-\mathcal{S}^{\Lambda}[\phi] +J\cdot\phi}\,,
	\end{equation}
where the UV regulated bare action was introduced in \eqref{total-bare}.
We introduce an intermediate cutoff scale $k$ 
by re-expressing the propagator as:
	\begin{equation}
	\label{Delta-partition}
	\tilde{\Delta}^\Lambda=\Delta^\Lambda_k+\Delta^k\,,
	\end{equation}
where $\tilde{\Delta}^\Lambda$, $\Delta^\Lambda_k$ and $\Delta^k$ are defined in \eqref{DeltaTotalUV}, \eqref{DeltaUVIR} and \eqref{DeltaUV} respectively, and the split above follows from the sum rule relation \eqref{sum rule}. The partition function can identically be rewritten as\footnote{up to a constant of proportionality. We ignore these from now on.}
	\begin{equation}
	\label{mod.int}
	Z^\Lambda[J]=\int\!\! \mathcal{D}\phi_{>}\mathcal{D}\phi_{<}\,\mathrm{e}^{-\frac{1}{2}\phi_{>}\cdot\left(\Delta^\Lambda_k\right)^{\!-1}\!\!\cdot\phi_{>}-
	\frac{1}{2}\phi_{<}\cdot\left(\Delta^k\right)^{\!-1}\!\!\cdot\phi_{<}-\mathcal{S}^{\Lambda}[\phi_{>}+\phi_{<}]+J\cdot(\phi_{>}+\phi_{<})}\,.
	\end{equation}
To see that this is true perturbatively, note that as a consequence of the sum form of the interactions, every Feynman diagram constructed from \eqref{mod.int} now appears twice for every internal propagator it contains: once with $\tilde{\Delta}^\Lambda$ replaced by $\Delta^\Lambda_k$ and once with $\tilde{\Delta}^\Lambda$ replaced by $\Delta^k$. Thus for every propagator line, what actually counts is the sum, which however is just $\tilde{\Delta}^\Lambda$ again by \eqref{Delta-partition} \cite{Morris:1998}. To prove the identity non-perturbatively, make the change of variables to $\phi=\phi_{>}+\phi_{<}$, for example by eliminating $\phi_>$. Evidently in \eqref{mod.int}, the action then has only up to quadratic dependence on $\phi_<$. Making the change of variables $\phi_<=\phi'_<+(\Delta^k/\tilde{\Delta}^\Lambda_k)\cdot\phi$, and using \eqref{Delta-partition}, results in the partition function factorising into a decoupled Gaussian integral over $\phi'_<$ (the constant of proportionality) and \eqref{func.int}, as required \cite{Morris:1993}.

Clearly, $\phi_{>}$ and $\phi_{<}$ beg to be regarded as the modes with momenta above and below $k$ respectively. This distinction is however only precise in the limit that the cutoff functions $C^\Lambda_k$ and $C^k$ become sharp. In general, modes in $\phi_{>}$ with $|p|<k$ and those in $\phi_{<}$ with $|p|>k$ will only be damped by the relevant cutoff functions. Even so, from now on we refer to $\phi_{>}$ ($\phi_{<}$) as high (low) momentum modes.

Consider computing the integral over the high momentum modes only in \eqref{mod.int}:
	\begin{equation}
	\label{high.int}
	Z^\Lambda_k[J, \phi_{<}]\equiv\int\!\! \mathcal{D}\phi_{>}\,\mathrm{e}^{-\frac{1}{2}\phi_{>}\cdot\left(\Delta^\Lambda_k\right)^{\!-1}\!\!\cdot\phi_{>}
	-\mathcal{S}^{\Lambda}[\phi_{>}+\phi_{<}]+J\cdot(\phi_{>}+\phi_{<})}
	\end{equation}
where $\phi_{<}$ now plays the r\^ole of a background field. Indeed, setting $\phi_{<}=0$ gives back the standard construction from which we can define the (UV and IR regulated) Legendre effective action,  \aka effective average action, as we will recall later:
	\begin{equation}
	\label{high.int-standard}
	Z^\Lambda_k[J]:=Z^\Lambda_k[J, 0]\equiv\int\!\! \mathcal{D}\phi_{>}\,\mathrm{e}^{-\frac{1}{2}\phi_{>}\cdot\left(\Delta^\Lambda_k\right)^{\!-1}\!\!\cdot\phi_{>}
	-\mathcal{S}^{\Lambda}[\phi_{>}]+J\cdot\phi_{>}}\,.
	\end{equation}
From \eqref{high.int}, performing again the linear shift $\phi_{>}=\phi-\phi_{<}$ and rewriting the interaction $\mathcal{S}^{\Lambda}$ as a function of $\delta/\delta J$ gives
	\begin{equation}
	Z^\Lambda_k[J, \phi_{<}]=\mathrm{e}^{-\frac{1}{2}\phi_{<}\cdot\left(\Delta^\Lambda_k\right)^{\!-1}\!\!\cdot\phi_{<}}\,\mathrm{e}^{-\mathcal{S}^{\Lambda}			[\frac{\delta}{\delta J}]}\int\!\!\mathcal{D}\phi \,\mathrm{e}^{-\frac{1}{2}\phi\cdot\left(\Delta^\Lambda_k\right)^{\!-1}\!\!\cdot\phi+
	\phi\cdot(J+\left(\Delta^\Lambda_k\right)^{\!-1} \cdot	\phi_{<})}\,.
	\end{equation}
Following another change of variables $\phi'=\phi-\Delta^\Lambda_k\cdot J-\phi_{<}$, the remaining integral is a decoupled Gaussian in $\phi'$ and, after some rearranging, we obtain
	\begin{align}
	Z^\Lambda_k[J, \phi_{<}]= &\,\mathrm{e}^{\frac{1}{2}J\cdot\Delta^\Lambda_k\cdot J+J\cdot\phi_{<}}\,\mathrm{e}^{-\frac{1}{2}(J+
	\left(\Delta^\Lambda_k\right)^{\!-1}\!\!\cdot\phi_{<})\cdot{\Delta^\Lambda_k}\cdot(J+\left(\Delta^\Lambda_k\right)^{\!-1}\!\!\cdot\phi_{<})}\times\nonumber\\
	&\quad\mathrm{e}^{-\mathcal{S}^{\Lambda}[\frac{\delta}{\delta J}]}\,\mathrm{e}^{\frac{1}{2}(J+
	\left(\Delta^\Lambda_k\right)^{\!-1}\!\!\cdot\phi_{<})\cdot{\Delta^\Lambda_k}\cdot(J+\left(\Delta^\Lambda_k\right)^{\!-1}\!\!\cdot\phi_{<})}\,.
	\end{align}
Performing all derivatives in $\mathcal{S}^{\Lambda}[\delta/\delta J]$, we find
	\begin{equation}
	\label{high.Z}
	Z^\Lambda_k[J,\phi_{<}]=\,\mathrm{e}^{\frac{1}{2}J\cdot{\Delta^\Lambda_k}\cdot J+J\cdot\phi_{<}-S^k[{\Delta^\Lambda_k}\cdot J+\phi_{<}]}
	\end{equation}
for some functional $S^k$. Substituting the above expression into \eqref{mod.int}, we have another identity \cite{Morris:1993} for the original partition function \eqref{func.int}:
	\begin{equation}
	\label{low.int}
	Z^\Lambda[J]=\int\!\!\mathcal{D}\phi_{<}\,\,\mathrm{e}^{-\frac{1}{2}\phi_{<}\cdot \left(\Delta^k\right)^{\!-1}\!\!\cdot \phi_{<}+
	\frac{1}{2}J\cdot {\Delta^\Lambda_k}\cdot J+J\cdot \phi_{<}-S^k[{\Delta^\Lambda_k}\cdot J+\phi_{<}]}\,.
	\end{equation}
All the high modes have been integrated out. Consider for the moment the case where $J$ couples only to low energy modes \ie so that ${\Delta^\Lambda_k}\cdot J=0$. Such is the case for example if the cutoff is of compact support so that $C^\Lambda_k(p)=0$ for $|p|<k$, and we choose $J$ to vanish for high energy modes, \ie $J(p)=0$ for $|p|>k$. Choosing $J(p)=0$ for $|p|>k$ of course just means not considering Green's functions with momenta greater than this effective cutoff. Then $Z^\Lambda[J]$ simplifies to
	\begin{equation}
	\label{low.int.simplified}
	Z^\Lambda[J]=\int\!\!\mathcal{D}\phi_{<}\,\,\mathrm{e}^{-\frac{1}{2}\phi_{<}\cdot \left(\Delta^k\right)^{\!-1}\!\!\cdot \phi_{<}-S^k[\phi_{<}]
	+J\cdot \phi_{<}}\,.
	\end{equation}
It is now straightforward to recognize the functional $S^k$ as the interaction part of the total Wilsonian effective action 
\eqref{total-Wilsonian}
regulated in the UV at $k$. 

Since \eqref{low.int} is nothing but the original partition function \eqref{func.int}, it gives Green's functions which are all actually independent of $k$, despite appearances to the contrary. Therefore, as advertised, $S^k$ plays the r\^ole of (the interactions in) a perfect action \cite{Hasenfratz:1993sp}. Note also that from \eqref{high.Z} and \eqref{high.int}, we obtain a prescription for computing the Wilsonian effective action from the bare action via a functional integral. We will return to this in sec. \ref{sec:solving}. 

The identification as a Wilsonian (perfect) action, is still valid if we let $J$ couple to all modes. We just have to recognise that it then also enters non-linearly with the precise prescription given in \eqref{low.int}, \ie as well as being the source it also plays the part of a space-time dependent coupling. Alternatively, we can use \eqref{low.int.simplified} even if ${\Delta^\Lambda_k}\cdot J\ne0$. In this case it is no longer true that \eqref{low.int.simplified} is independent of $k$, since we are missing the terms in \eqref{low.int} that contribute to making $Z^\Lambda[J]$ and thus all Green's functions independent of $k$. However for Green's functions all of whose (external) momenta $|p|\ll k$, we have $\Delta^\Lambda_k(p)=0$ to very good approximation. Furthermore $\Delta^\Lambda_k(p)\to0$ as $|p|/k\to0$, implying that in this limit \eqref{low.int.simplified} becomes exactly independent of $k$.

The flow equation for $S^k$ is found by first differentiating \eqref{high.int} with respect to $k$ to obtain the flow equation for $Z^\Lambda_k[J,\phi_{<}]$:
	\begin{equation}
	\label{flow.Z}
	\frac{\partial}{\partial k}Z^\Lambda_k[J,\phi_{<}]=-\frac{1}{2}\bigg(\frac{\delta}{\delta J}-\phi_{<}\bigg)\cdot						\bigg(\frac{\partial}{\partial k}\left(\Delta^\Lambda_k\right)^{\!-1}\bigg)\cdot \bigg(\frac{\delta}{\delta J}-\phi_{<}\bigg)Z^\Lambda_k[J,\phi_{<}]\,.
	\end{equation}
Then by inserting \eqref{high.Z} into the above expression 
and defining $\Phi\equiv{\Delta^\Lambda_k}\cdot J+\phi_{<}$, we obtain
exactly the already advertised Polchinski flow equation \eqref{S.flow}.

Turning our attention to \eqref{high.Z} once more, we can recognise it as being related to the generator of connected Green's functions $W^\Lambda_k$ with IR cutoff $k$:
	\begin{equation}
	\label{W}
	\mathrm{e}^{W^\Lambda_k[J,\phi_{<}]}\equiv Z^\Lambda_k[J,\phi_{<}]=\,\mathrm{e}^{\frac{1}{2}J\cdot {\Delta^\Lambda_k}\cdot J+J\cdot \phi_{<}
	-S^k[{\Delta^\Lambda_k}\cdot J+\phi_{<}]}
	\end{equation}
and in taking the limit $k\rightarrow 0$, we recover the standard Green's functions (regulated in the UV through $\tilde{\Delta}^\Lambda$). The Legendre transform of $W^\Lambda_k$ gives the  Legendre effective action $\Gamma^{\text{tot}}_{k}$: 
	\begin{align}
	\label{Legendre}
	\Gamma^{\text{tot},\Lambda}_{k}[\varphi,\phi_{<}]&=-W^\Lambda_k[J,\phi_{<}]+J\cdot \varphi\\ \label{Legendre-interactions}
	&=\frac{1}{2}(\varphi-\phi_{<})\cdot \left(\Delta^\Lambda_k\right)^{\!-1}\!\!\cdot (\varphi-\phi_{<})+\Gamma^\Lambda_k[\varphi]
	\end{align}
where $\varphi\equiv\delta W^\Lambda_k/\delta J$ is the classical field and $\Gamma^\Lambda_k$ is the interaction part which carries no $\phi_{<}$ dependence \cite{Morris:1993}, as follows from 
\be 
\label{no-phi<}
\frac{\delta}{\delta\phi_<}\Gamma^{\text{tot},\Lambda}_{k}[\varphi,\phi_{<}] = -\frac{\delta}{\delta\phi_<}W^\Lambda_k[J,\phi_{<}] 
 =-\left(\Delta^\Lambda_k\right)^{\!-1}\!\!\cdot\left(\frac{\delta W^\Lambda_k}{\delta J}-\phi_<\right) = \left(\Delta^\Lambda_k\right)^{\!-1}\!\!\cdot\left(\phi_<-\varphi\right)\,,
\ee
where we have used \eqref{Legendre} and then \eqref{W}. 

Notice that when $\phi_<=0$, we have the standard definition of the partition function \eqref{high.int-standard} and from it the standard definition of $W^\Lambda_k[J]$ in \eqref{W} and thus from \eqref{Legendre} the standard definition of the (IR and UV regulated) Legendre effective action. Thus from \eqref{Legendre-interactions} with $\phi_<=0$,   it follows that 
$\Gamma^\Lambda_k[\varphi]$ is the same interactions part of the effective average action as defined in \eqref{total-Gamma}. See also the discussion in sec. \ref{sec:details} leading up to  \eqref{total-Gamma}.  Recall that $\Gamma^\Lambda_k[\varphi]$ is thus equivalently the interactions part of the generator of one particle irreducible (1PI) Green's functions, cutoff in the IR at $k$, and coincides with the interactions part of the standard effective action $\Gamma$ in the limit $k\rightarrow0$. 
 
 Substituting the Legendre transform equation \eqref{Legendre} into \eqref{flow.Z}, we obtain the already advertised flow equation \eqref{Gamma.flow} for $\Gamma^\Lambda_k$. From equation \eqref{Legendre} follows the well known fact that connected Green's functions can be expressed as a tree level sum of 1PI vertices (in this case connected by IR cutoff propagators). Thus equation \eqref{W} implies that the vertices of $S^k$ will also have a similar expansion (see section \ref{sec:vertices}). Indeed, we can find a direct relationship between $S^k$ and $\Gamma^\Lambda_k$ by substituting \eqref{W} into \eqref{Legendre}, using \eqref{Legendre-interactions} and recalling that $\Phi={\Delta^\Lambda_k}\cdot J+\phi_{<}$. The result is the duality equation \eqref{duality-gen} we have been aiming for.
 
To reiterate, \eqref{duality-gen} is an exact relationship between the interaction part of the Wilsonian effective action, $S^k$, regulated in the UV at $k$ and the interaction part of the Legendre effective action, $\Gamma^\Lambda_k$ regulated in the UV at $\Lambda$ and regulated in the IR at $k$ (\aka effective average action). It gives rise to a duality between the flow equations \eqref{S.flow} and \eqref{Gamma.flow}. If we have a complete RG trajectory for $\Gamma_k$, that is a solution of \eqref{R-flow} where the UV cutoff $\Lambda$ has been removed, and where by complete we mean that it extends from a UV fixed point as $k\rightarrow \infty$ down to $k\rightarrow 0$, then we can take the continuum limit of the key equations given in this section simply by replacing $\Delta^\Lambda_k$ with $\Delta_k$. In this way we equivalently have a solution to \eqref{Gamma.flow}, the duality relation now reads \eqref{duality-con}, which allows us to compute the equivalent RG trajectory for $S^k$ with the equivalent fixed point solution, and the continuum limit of the effective partition functions can then be computed directly from \eqref{low.int} and \eqref{low.int.simplified}.

\section{Vertices of the Wilsonian Effective Action}
\label{sec:vertices}

In this section we use result \eqref{duality-con} to derive explicit expressions for the vertices of $S^k$ in terms of those of $\Gamma_{k}$. Clearly \eqref{duality-con} is symmetric under the map: $S_k\leftrightarrow\Gamma_k$ with $\Delta_k\mapsto-\Delta_k$, so by relabelling in this way we can also use the expressions below to derive the vertices of $\Gamma_k$ from $S^k$.
Clearly these expressions can therefore also be used after some renaming to give the vertices of one action in terms of another for any of the alternative expressions of duality, namely  \eqref{duality-gen},  \eqref{duality-Gamma-con}, \eqref{duality-gen-2} and \eqref{duality-Gamma-gen}. For example to obtain the vertices of $\Gamma^\Lambda_k$ in terms of those of $\Gamma_k$ using \eqref{duality-Gamma-con} (part of our second solution to the reconstruction problem) it is only necessary to replace $S^k$ with $\Gamma^\Lambda_k$ and $\Delta_k$ with $\Delta_\Lambda$ in the following expressions.

Extracting the momentum conserving Dirac delta-function in what follows, vertices of $S^k$ will be denoted by
	\begin{equation}
	(2\pi)^d\delta(p_1+\cdots+p_n)\,S^{(n)}(p_{1},\cdots,p_{n};k)\equiv\frac{\delta^{n} S^k[\Phi]}{\delta\Phi(p_{1})\cdots\delta\Phi(p_{n})}\bigg|				_{\Phi=0}
	\end{equation}
and the vertices of $\Gamma_{k}$ by
	\begin{equation}
	\label{Gamma-vertices}
	(2\pi)^d\delta(p_1+\cdots+p_n)\,\Gamma^{(n)}(p_{1},\cdots,p_{n};k)\equiv\frac{\delta^{n} \Gamma_{k}[\Phi]}{\delta\Phi(p_{1})\cdots						\delta\Phi(p_{n})}\bigg|_{\Phi=0}
	\end{equation}
with the exception of its 2-point function which we write as $\Sigma(p^{2};k)$. We often omit the momentum arguments of the vertices for neatness. For simplicity we impose a $Z_{2}$ symmetry $\phi \leftrightarrow -\phi$ on ${\cal S}^\Lambda$ so that it only contains even powers of $\phi$ and hence $S^{(n)}(p_{1},\cdots,p_{n};k)$ and $\Gamma^{(n)}(p_{1},\cdots,p_{n};k)$ vanish for odd $n$.

We start by writing \eqref{duality-con} more conveniently as
	\begin{equation}
	\label{soln2}
	S^k[\Phi]=\Gamma_{k}[\Phi-{\Delta_k}\cdot \frac{\delta S^k}{\delta\Phi}]+\frac{1}{2}\frac{\delta S^k}				{\delta\Phi}\cdot {\Delta_k}\cdot \frac{\delta S^k}{\delta\Phi}
	\end{equation}
by recognising that $\varphi=\Phi-{\Delta_k}\cdot (\delta S^k/\delta\Phi)$.  Taylor expanding the right hand side, keeping only bilinear terms in $\Phi$ and rearranging, we find the following expression for the 2-point function:
	\begin{equation}
	\label{two.point}
	S^{(2)}(p^{2};k)=\Sigma(p^{2};k)\left(1+{\Delta_k(p)}\Sigma(p^{2};k)\right)^{\!-1}\,.
	\end{equation}
Expanding the RHS perturbatively in $\Sigma$ gives the expected expansion of $S^{(2)}$ in terms of 1PI vertices, connected by IR cutoff propagators. Note that in obtaining this result, it is only necessary to expand to second order in the Taylor series as the $Z_{2}$ symmetry kills the cross-terms from one-point and three-point vertices that would otherwise appear.

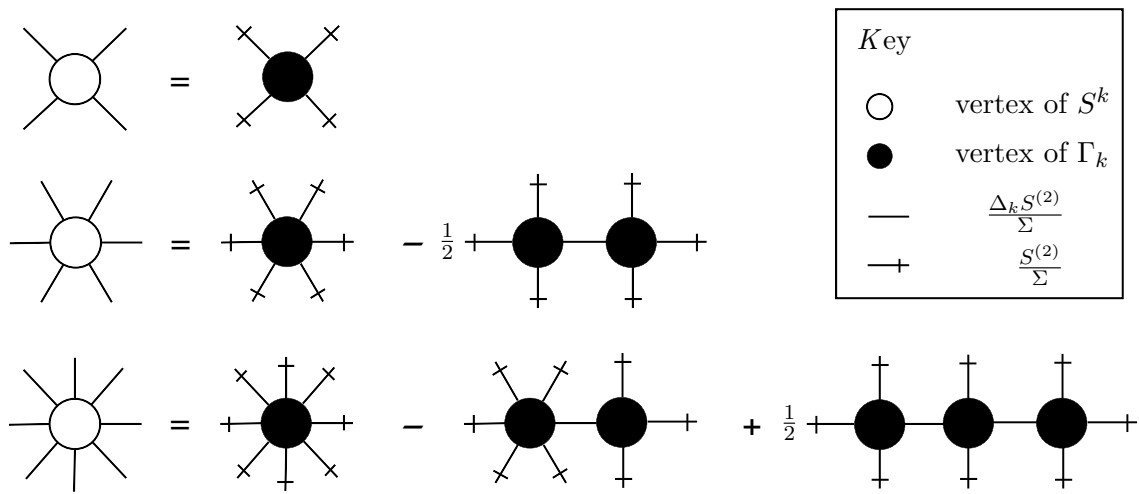
\begin{figure}
	\begin{center}
	\setlength{\unitlength}{2mm}
	\begin{picture}(75,35)
	\includegraphics[scale=0.9]{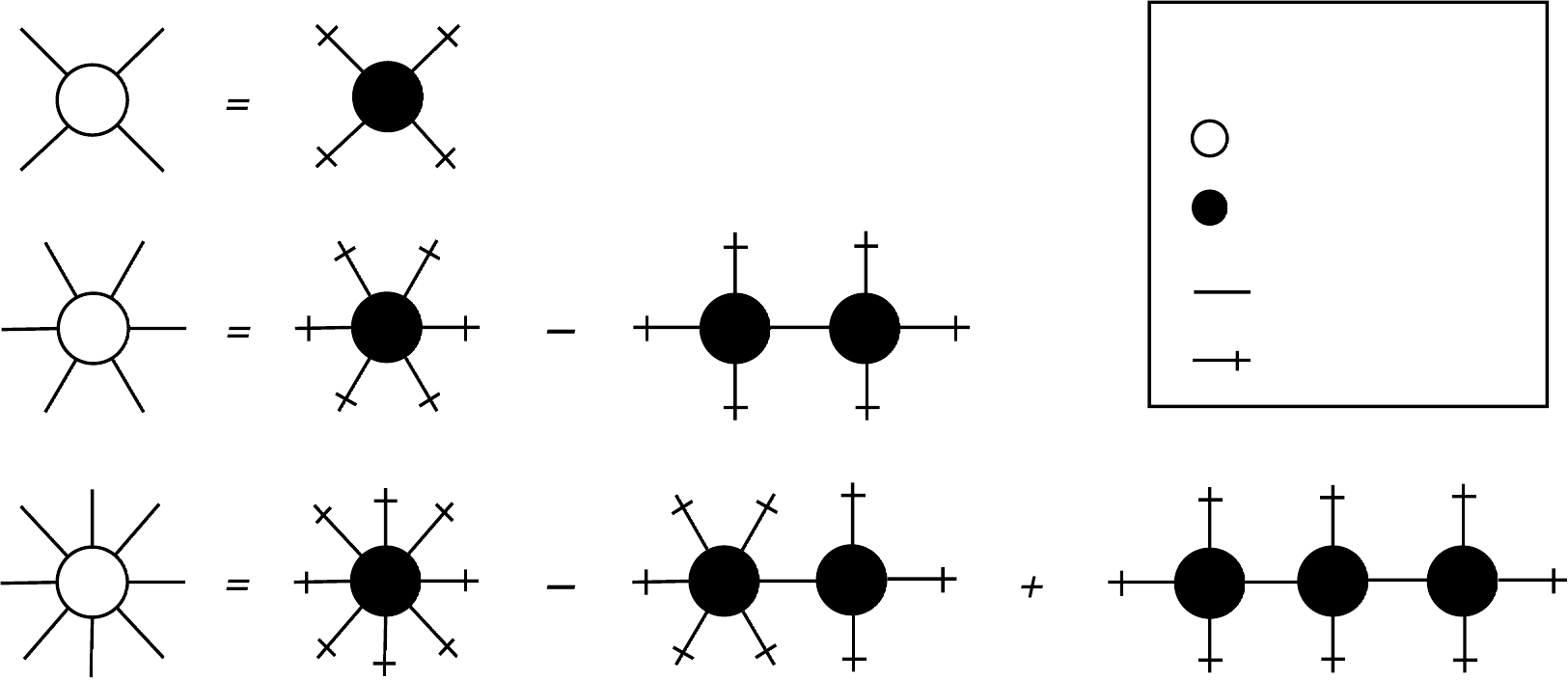}
	\put(-18.5,29.5){\emph Key}
	\put(-12,25.25){vertex of $S^k$}
	\put(-12, 22){vertex of $\Gamma_{k}$}
	\put(-10,18){$\frac{\Delta_kS^{(2)}}{\Sigma}$}
	\put(-8, 14.5){$\frac{S^{(2)}}{\Sigma}$}
	\put(-46, 16.25){$\frac{1}{2}$}
	\put(-23.5,4.2){$\frac{1}{2}$}
	\end{picture}
	\end{center}
	\caption{Vertices of the Wilsonian effective interaction $S^k$ for $n=4,6$ and 8 respectively. Each diagram containing more than one vertex represents a sum over disjoint subsets of momenta corresponding to the number and type of vertices in each diagram {\it e.g.} the final diagram in the expansion of $S^{(8)}$ stands for a sum over partitions of $\{p_{1},\cdots,p_{8}\}$ into 2 sets of 3 momenta and 1 set of 2 momenta.\label{fig:vertices}}
\end{figure}

To compute expressions for vertices for $n>2$, we need to isolate the 2-point pieces from $S^k$ and $\Gamma_{k}$, like so
	\begin{align}
	S^k[\Phi]=\frac{1}{2}\Phi\cdot S^{(2)}\cdot \Phi + S'^k[\Phi]&&
	\Gamma_{k}[\varphi]=\frac{1}{2}\varphi\cdot \Sigma\cdot \varphi + \Gamma'_{k}[\varphi]
	\end{align}
such that all terms but those quadratic in the fields are contained in $S'^k$ and $\Gamma'_{k}$. Upon substituting the above into \eqref{soln2} and using \eqref{two.point}, we have
	\begin{equation}
	\label{soln3}
	S'^k[\Phi]=\Gamma_{k}'[\frac{S^{(2)}}{\Sigma}\cdot \Phi-{\Delta_k}\cdot \frac{\delta S'^k}{\delta\Phi}]+\frac{1}				{2}\frac{\delta S'^k}{\delta\Phi}\cdot \frac{\Delta_k\Sigma}{S^{(2)}}\cdot \frac{\delta S'^k}{\delta\Phi}\,.
	\end{equation}
Again, by Taylor expanding the right-hand side to the desired order, we obtain our vertex of choice. In general, for an $n$-point function, we only have to keep terms in the Taylor series up to and including the $n$th order: higher order terms vanish either from the $Z_{2}$ symmetry or because they then contain too many $\Phi$s. For the 4-point function we have
	\begin{equation}
	\label{4.point}
	S^{(4)}(p_{1},p_{2},p_{3},p_{4};k)=\Gamma^{(4)}(p_{1},p_{2},p_{3},p_{4};k)\prod_{i=1}^{4}\frac{S^{(2)}			(p^{2}_{i};k)}{\Sigma(p^{2}_{i};k)}\,.
	\end{equation}
Likewise, the 6-point function is given by
	\begin{align}
	S^{(6)}(p_{1},\cdots,p_{6};k)=&\Gamma^{(6)}(p_{1},\cdots,p_{6};k)\prod_{i=1}^{6}\frac{S^{(2)}(p^{2}_{i};k)}				{\Sigma(p^{2}_{i};k)}\nonumber\\
	&\quad-\frac{1}{2}\sum_{\{I_{1},I_{2}\}}\bigg\{\Gamma^{(4)}(I_{1},q;k)\prod_{p_{i}\in I_{1}}
	\frac{S^{(2)}(p_{i}^{2};k)}{\Sigma(p_{i}^{2};k)}\nonumber\\
	&\qquad\times{\Delta_k}(q^{2})\frac{S^{(2)}(q^{2};k)}{\Sigma(q^{2};k)}\Gamma^{(4)}(-q,I_{2};k)\prod_{p_{j}
	 \in I_{2}}\frac{S^{(2)}(p_{j}^{2};k)}{\Sigma(p_{j}^{2};k)}\bigg\}
	\end{align}
where $I_{1}$ and $I_{2}$ are disjoint subsets of 3 momenta such that $I_{1}\cup I_{2}=\{p_{1},\cdots,p_{6}\}$. The sum over $\{I_{1},I_{2}\}$ means sum over all such subsets. By momentum conservation, the momentum $q$ carried by certain 2-point functions is equivalent to a partial sum {\it i.e.} $q=\sum_{p_{i} \in I} p_{i}$ where $I$ is a subset of the total set of external momenta. Graphical representations of these expressions, as well as one for the 8-point function, are given in figure \ref{fig:vertices} and are much easier to interpret. Of course the expansion can be continued to higher orders.

\section{Second solution to the reconstruction problem}
\label{sec:solving}

In this section we provide more detail on our second solution to the reconstruction problem and show how it is related to the one-loop approximate solution \eqref{MandR.soln} provided in ref. \cite{Manrique:2008zw}. As explained in the Introduction and sec. \ref{sec:details}, given a complete RG trajectory for $\Gamma_k[\varphi]$, \eqref{duality-con} then provides us with $S^k[\Phi]$ which is the interaction part of a perfect bare action. This already provides us with an acceptable solution to the reconstruction problem, but as we emphasised in appendix \ref{app:UV} it cannot give us back $\Gamma_k$ via the standard path integral route \eqref{high.int-standard} since such a UV regulated path integral necessarily leaves its imprint on the Legendre effective action such that it now depends on both cutoffs: $\Gamma\equiv\Gamma^\Lambda_k$. However what can be done is to use $\Gamma_k[\varphi]$ to construct a pair 
$\{\bS, \Gamma_{k}^\Lambda\}$, where $\Gamma^\Lambda_k$ is related to $\bS$ in the usual way, and such that as $\Lambda\to\infty$ we have $\Gamma_{k}^\Lambda\to\Gamma_k$. This is our second solution. The question then is how this solution is to be compared with the one-loop approximate relation \eqref{MandR.soln}.

Let us first note that in \eqref{MandR.soln} we can split off the bare interactions and effective average action interactions as defined in \eqref{other-interactions} and \eqref{interactions} respectively. For the left hand side of \eqref{MandR.soln} that just means dropping the hats, 
but in the right hand side we recognise that as in the shift from \eqref{LEAA-con} to \eqref{LEAA-alt} we can incorporate the infrared cutoff through a multiplicative profile \eqref{IR-only} and then make explicit the UV sharp cutoff by replacing this by \eqref{UVIRMR}. The net result is that we re-express equation \eqref{MandR.soln} as
\begin{equation}
	\label{descents.soln.1}
	\Gamma_{k=\Lambda}^\Lambda[\varphi]=\mathcal{S}^{\Lambda}[\varphi]+
	\frac{1}{2}\text{tr}\,\text{ln}\Big\{\mathcal{S}^{\Lambda (2)}[\varphi]+\left(\Delta^\Lambda_\Lambda\right)^{\!-1}\Big\}\,.
	\end{equation}
This has two advantages. Firstly it makes the overall UV sharp cutoff explicit, and secondly actually this formula is valid as a one-loop approximation in general, whatever the precise profile of IR and UV cutoff we implement via $C^\Lambda_k(p)$. 
The total effective average action is then in general given as in \eqref{total-Gamma} and the total bare action as in \eqref{total-bare-k}. As already reviewed below \eqref{no-phi<}, the standard construction using the partition function \eqref{high.int-standard} yields of course this $\Gamma^\Lambda_k[\ph]$. 

Our second solution to the reconstruction problem follows  from employing compatible cutoffs. As defined in the Introduction, by compatible cutoffs we mean that $C^\Lambda_\Lambda(p)=0$ for all $p$, \ie such that when the IR cutoff meets the UV cutoff the result completely kills the propagator: $\Delta^\Lambda_\Lambda \equiv 0$. Up to a (divergent but irrelevant) constant we then have $\Gamma_{k=\Lambda}^\Lambda[\varphi]=\mathcal{S}^{\Lambda}[\varphi]$ as is clear from \eqref{descents.soln.1} if we note that
\be 
\label{one-loop-compatible}
\text{tr}\,\text{ln}\Big\{\mathcal{S}^{\Lambda (2)}[\varphi]+\left(\Delta^\Lambda_\Lambda\right)^{\!-1}\Big\} = -\text{tr}\,\text{ln}\left\{\Delta^\Lambda_\Lambda\right\} +\text{tr}\,\text{ln}\Big\{1+\Delta^\Lambda_\Lambda\cdot\mathcal{S}^{\Lambda (2)}[\varphi]\Big\}\,.
\ee
Indeed the fact that in \eqref{high.int-standard}, $\left(\Delta^\Lambda_k\right)^{\!-1}\to\infty$ as $k\to\Lambda$, turns the steepest descents calculation that gives \eqref{descents.soln.1} into an exact statement. For the same reason, from the most direct expression relating the Wilsonian interactions $S^k$ to the bare interactions, obtained by setting $J=0$ in \eqref{high.int} and \eqref{high.Z}:
	\begin{equation}
	\label{partition to S}
	Z^\Lambda_k[0,\phi_{<}]=\,\mathrm{e}^{-S^k[\phi_{<}]}=\int\!\!\mathcal{D}\phi_{>}
	\,\,\mathrm{e}^{-\frac{1}{2}\phi_{>}\cdot \left(\Delta^\Lambda_k\right)^{\!-1}\!\!\cdot \phi_{>}-
	\mathcal{S}^{\Lambda}[\phi_{>}+\phi_{<}]}\,,
	\end{equation}
we see that we have no choice but to have the equality $S^{\Lambda}[\varphi]=\mathcal{S}^{\Lambda}[\varphi] $. To make the map from the continuum solution $\Gamma_k$ to this system, we insist that the Wilsonian interactions $S^k$ and thus also the effective Wilsonian cutoff $C^k(p)$, are still  the continuum ones. Then this fixes via \eqref{sum rule} the overall bare cutoff to be the continuum Wilsonian one: $\C=C^\Lambda$, and as we see already the bare interactions must taken to be $\mathcal{S}^{\Lambda}[\varphi] = S^{\Lambda}[\varphi]$. Then the map \eqref{duality-Gamma-con} from $\Gamma_k$ to $\Gamma^\Lambda_k$ follows, as proved in sec. \ref{sec:duality-Gamma}, and worked out in detail in sec. \ref{sec:vertices}. We thus have all the elements of our second solution.

If the UV and IR cutoff imposed by ${\Delta^\Lambda_k}$ are not compatible, then $\Delta^\Lambda_\Lambda\ne0$ and in both \eqref{high.int-standard} and \eqref{partition to S} there is still a non-trivial functional integral to compute in the limit $k\to\Lambda$. To one loop, the result for $\Gamma^\Lambda_\Lambda$ is the one given in \eqref{descents.soln.1}.
 In analogy with \cite{Manrique:2008zw}, let us also compute the integral in \eqref{partition to S} to one loop, using the method of steepest descents. The exponent is at a minimum when
	\begin{equation}
	\label{min}
	\phi_{>}=-{\Delta^\Lambda_k}\cdot \frac{\delta \mathcal{S}^{\Lambda}[\phi_>+\phi_<]}{\delta\phi_{>}}\equiv\phi^{0}_{>}\,.
	\end{equation}
We define $\phi_{>}\equiv\phi_{>}^{0}+\tilde{\phi}_{>}$ 
and expand about $\tilde{\phi}_{>}=0$, keeping only up to second derivatives of $\mathcal{S}^{\Lambda}$:
	\begin{equation}
	\mathrm{e}^{-S^k[\phi_{<}]}=\,\mathrm{e}^{-\frac{1}{2}\phi^{0}_{>}\cdot \left(\Delta^\Lambda_k\right)^{\!-1}\!\!\cdot \phi^{0}_{>}}
	\,\mathrm{e}^{-\mathcal{S}^{\Lambda}[\phi_{>}^{0}+\phi_{<}]}
	\int\!\!\mathcal{D}\tilde{\phi}_{>}\,\,\mathrm{e}^{-\frac{1}{2}\tilde{\phi}_{>}\cdot \left(\Delta^\Lambda_k\right)^{\!-1}\!\!\cdot \tilde{\phi}_{>}
	-\frac{1}{2}\tilde{\phi}_{>}\cdot \frac{\delta^{2}\mathcal{S}^{\Lambda}}{\delta\phi_{>}\delta\phi_{>}}\cdot \tilde{\phi}_{>}}\,.
	\end{equation}	
The terms linear in $\tilde{\phi}_{>}$ cancel by \eqref{min}. Performing the Gaussian integral over $\tilde{\phi}_{>}$, we find
	\begin{equation}
	\label{one-loop-Wilsonian}
	S^k[\phi_{<}]-\frac{1}{2}\phi_{>}^{0}\cdot \left(\Delta^\Lambda_k\right)^{\!-1}\!\!\cdot \phi_{>}^{0}
	=\mathcal{S}^{\Lambda}[\phi^{0}_{>}\!+\!\phi_{<}]+
	\frac{1}{2}\text{tr}\,\text{ln}
	\Big\{\frac{\delta^{2}\mathcal{S}^{\Lambda}[\phi^{0}_{>}\!+\!\phi_{<}]}									{\delta\phi_{>}\delta\phi_{>}}+\left(\Delta^\Lambda_k\right)^{\!-1}\Big\}\,.
	\end{equation}
Introducing $\varphi\equiv\phi_{>}^{0}+\phi_{<}$, we thus have
	\begin{equation}
	\label{descents.soln-S}
	S^k[\phi_{<}]-\frac{1}{2}(\varphi-\phi_<)\cdot \left(\Delta^\Lambda_k\right)^{\!-1}\!\!\cdot (\varphi-\phi_<)=\mathcal{S}^{\Lambda}[\varphi]+
	\frac{1}{2}\text{tr}\,\text{ln}\Big\{\mathcal{S}^{\Lambda (2)}[\varphi]+\left(\Delta^\Lambda_k\right)^{\!-1}\Big\}\,.
	\end{equation}
Comparing \eqref{descents.soln.1} we recognise that the right hand side is nothing but the one-loop approximation to the effective average action at a general value of $k$:
	\begin{equation}
	\label{descents.soln}
	\Gamma_{k}^\Lambda[\varphi]=\mathcal{S}^{\Lambda}[\varphi]+
	\frac{1}{2}\text{tr}\,\text{ln}\Big\{\mathcal{S}^{\Lambda (2)}[\varphi]+\left(\Delta^\Lambda_k\right)^{\!-1}\Big\}\,.
	\end{equation}
Finally comparing \eqref{descents.soln} and \eqref{descents.soln-S}, we see that we recover  the duality relation \eqref{duality-gen} in sec. \ref{sec:SandGamma}.\footnote{It can also be shown that this is consistent to one loop with the solution \eqref{min}.} We have therefore explicitly confirmed the duality relation to one loop via the steepest descents method.
Through the above demonstration and also our discussion of the compatible case, \cf \eqref{one-loop-compatible}, we have also comprehensively explored how our solution is related to the one-loop result \eqref{MandR.soln}.

\section{Some compatible cutoffs}
\label{sec:compatible}

In this section we briefly explore some possible forms of compatible cutoffs, \ie such that $C^\Lambda_k(p)$ vanishes identically when $k\to\Lambda$. We also insist that the effective Wilsonian UV cutoff $C^k(p)$ depends only on the one cutoff scale $k$ as indicated. Through the sum rule \eqref{sum rule} it follows that we take the overall UV cutoff to be the Wilsonian one at scale $\Lambda$: $\C=C^{k=\Lambda}$.

There are various possibilities for compatible cutoffs. One straightforward option is to make all the cutoff functions sharp:
	\begin{equation}
	\label{sharp}
	C^\Lambda= \left\{
  	\begin{array}{l l}
   	 0 & \quad |p|\geq\Lambda\vspace{3pt}\\
    	1 & \quad |p|<\Lambda\vspace{3pt}
 	 \end{array} 
 	 \right., \qquad
	C^k= \left\{
  	\begin{array}{l l}
   	 0 & \quad |p|>k\vspace{3pt}\\
    	 1& \quad  |p|\leq k\vspace{3pt}
 	 \end{array} 
 	 \right.,\qquad
	C^\Lambda_k= \left\{
  	 \begin{array}{l l}
   	 0 & \quad |p|\geq\Lambda \vspace{3pt}\\
    	 1 & \quad k<|p|<\Lambda\vspace{3pt}\\
	 0& \quad |p|\leq k\vspace{3pt}
 	 \end{array} 
 	 \right.\,.
	\end{equation}
Another choice of compatible cutoffs is:
	\begin{equation}
	\label{choice.2}
	C^\Lambda= \left\{
  	\begin{array}{l l}
   	 0 &  |p|\geq\Lambda\vspace{3pt}\\
    	1-\frac{p^{2}}{\Lambda^{2}} &  |p|<\Lambda\vspace{3pt}
 	 \end{array} 
 	 \right.,\quad
	C^k= \left\{
  	\begin{array}{l l}
   	 0 &  |p|\geq k\vspace{3pt}\\
    	 1-\frac{p^{2}}{k^{2}}&   |p|<k\vspace{3pt}
 	 \end{array} 
 	 \right.,\quad
	C^\Lambda_k=\left\{
  	\begin{array}{l l}
   	 0 &  |p|\geq\Lambda \vspace{3pt}\\
    	1-\frac{p^{2}}{\Lambda^{2}} &  k\leq |p|<\Lambda\vspace{3pt}\\
	\frac{p^{2}}{k^{2}}-\frac{p^{2}}{\Lambda^{2}}&  |p|<k\vspace{3pt}
 	 \end{array} 
 	 \right.\,.
 	\end{equation}
It can be easily checked that all cutoff functions have the desired regulating behaviour. Again, for $k=\Lambda$, we have ${\Delta^\Lambda_k}=0$. These cutoff functions have been obtained by, first of all, using \eqref{IR-only} to find the multiplicative IR cutoff function $C_k$ corresponding to the optimized cutoff. In order to ensure that the effective Wilsonian UV cutoff depends only on the one cutoff scale $k$, we   define it as $C^k=1-C_k$, \ie via \eqref{sum rule} but with the overall cutoff $\Lambda\to\infty$, and thus $\C\to1$.  (The result agrees with \eqref{MR-Wilson} since we already found that cutoff  $C^k$ to be dual to the optimised IR cutoff and also to be independent of $\Lambda$.) 
As we have seen, compatibility for finite overall cutoff then requires $\C\equiv C^\Lambda$. This however forces us to change the IR profile via \eqref{sum rule} to one, $C^\Lambda_k = C^\Lambda-C^k$, that includes both cutoffs. The resulting choices \eqref{choice.2} thus also have the property that as $\Lambda\to\infty$, $C^\Lambda_k$ returns to the (multiplicative form \eqref{IR-only} of the) optimised cutoff.

Another choice of additive IR regulator from which we can define compatible cutoff functions following these steps is
	\begin{equation}
	\label{alt-additive}
	\tilde{R}_{k}(p^{2})=\frac{1}{\mathrm{e}^{\frac{p^{2}}{k^{2}}}-1}\,.
	\end{equation}
This corresponds to the following choice of cutoffs:
	\begin{equation}
	C^k=\frac{1}{1+p^{2}\big(\mathrm{e}^{\frac{p^{2}}{k^{2}}}-1\big)}\,,
	\end{equation}
again the overall UV cutoff is just $C^\Lambda$, and
	\begin{equation}	
	C^\Lambda_k=\frac{p^{2}\big(\mathrm{e}^{\frac{p^{2}}{k^{2}}}-\mathrm{e}^{\frac{p^2}{\Lambda^2}}\big)}							{\big(1+p^2\big(\mathrm{e}^{\frac{p^2}{\Lambda^2}}-1\big)\big)\big(1+p^2\big(\mathrm{e}^{\frac{p^2}{k^2}}-1\big)\big)}\,.
	\end{equation}
These cutoffs regulate as required, exhibiting the behaviour described below \eqref{IR-only}, below \eqref{sharpUV} and below \eqref{MR-Wilson} respectively. Again we have defined the cutoffs so that  ${\Delta^\Lambda_k}=0$ when $k=\Lambda$,  whilst as $\Lambda\to\infty$, $C^\Lambda_k$ returns to the multiplicative version of \eqref{alt-additive}.
In summary, we have seen how we can formulate compatible cutoff functions
using a sharp cutoff, or based closely on the optimised cutoff $R_{k}$ or $\tilde{R}_{k}$.

\section{Conclusions}
\label{sec:conclusions}

Let us start by briefly summarising our main conclusions. We set out two solutions to the reconstruction problem, giving the recipes in detail in sec. \ref{sec:details}. Starting from a full renormalised trajectory for the effective average action \eqref{interactions}, whose interactions are given by $\Gamma_k[\varphi]$, we can reconstruct a suitable bare action by using the corresponding Wilsonian interactions $S^k[\Phi]$. This also describes the full renormalised trajectory, but in the Wilsonian language. $S^k[\Phi]$ is computed through the continuum duality relation \eqref{duality-con}. The vertices are then related via a tree expansion to the vertices of $\Gamma_k$ and these are worked out in detail in sec. \ref{sec:vertices}. 
The full Wilsonian effective action $S^{\mathrm{tot},k}[\Phi]$ is given by \eqref{total-Wilsonian}, where the effective multiplicative UV cutoff profile $C^k(p)=1-C_k(p)$, and $C_k$ is the multiplicative version of the additive IR cutoff via the translation \eqref{IR-only}. The partition function constructed using $S^{\mathrm{tot},k}[\Phi]$ is actually independent of $k$, and thus this bare action is an example of a perfect bare action. Written in the form \eqref{low.int.simplified} (where the superscript $\Lambda=\infty$ since we have taken the continuum limit), the independence with respect to $k$ is only approximate, becoming exact when we compute Green's functions with momenta $|p|\ll k$, unless the source $J$ obeys some restrictions, as discussed around \eqref{low.int.simplified}. Alternatively we can embed the source inside the action as well, as in \eqref{low.int}, and then the independence with respect to $k$ is indeed exact. 

A potential problem with this first solution to the reconstruction problem is that we have only the one cutoff $k$ involved which now plays the r\^ole of a UV cutoff for this perfect bare action. For some purposes we may want to investigate a system where a suitable bare action with UV regularisation set at some scale $\Lambda$ gives back the effective average action through the usual procedure. In other words, we insert an infrared cutoff $k$ into the bare action to give \eqref{total-bare-k}, where the overall multiplicative UV cutoff has been replaced by $C^\Lambda_k$ incorporating also the IR cutoff, and then form the partition function \eqref{high.int-standard}. As we emphasised in appendix \ref{app:UV}, we cannot get the continuum $\Gamma_k$ in such a way, since it is then guaranteed that the effective average action $\hat{\Gamma}^\Lambda_k$, bilinear part and interactions, now depends on both cutoffs, as displayed in \eqref{total-Gamma}.  What we can do however is again to take the bare interactions to be the perfect Wilsonian ones computed from $\Gamma_k$, thus $\bS = S^{k=\Lambda}$, and then the above procedure gives us a $\Gamma^\Lambda_k[\varphi]$, such that as $\Lambda\to\infty$, $\Gamma^\Lambda_k\to\Gamma_k$. The UV boundary conditions on the flow equation \eqref{Gamma.flow} for this effective average action are just $\Gamma^\Lambda_{k=\Lambda} = S^\Lambda = \bS$. 
We do not need to compute the functional integral, or the flow equation, to find $\Gamma^\Lambda_k[\varphi]$ however, since it is also directly related to the original continuum $\Gamma_k$ via a duality relation \eqref{duality-Gamma-con}, which may also be solved vertex by vertex as in sec. \ref{sec:vertices}. This is our second solution to the reconstruction problem.

We proved the latter duality relation by first proving an even more remarkable duality relation in sec. \ref{sec:duality-Gamma}, namely \eqref{duality-Gamma-gen}. This is a tree-level relation between two effective average actions computed with different overall cutoff profiles $\C$ and $\Cp$, but whose corresponding effective Wilsonian actions $S^{\mathrm{tot},k}$ actually coincide.  As we explain in sec. \ref{sec:duality-Gamma}, this assumes that the bare interactions $\bS$ and $\mathring{{\cal S}}^\Lp$ can be chosen precisely to ensure this. If we choose a solution $S^k$ of its flow equation \eqref{S.flow} that does not correspond to a full renormalised trajectory, then clearly this is not always possible, for example it is then not possible to raise the overall cutoff $\Lambda$ or $\Lp$ all the way to infinity. Even if we choose $S^k$ to be a renormalised trajectory, it still may not be possible to change the bare cutoff arbitrarily in such a way. The ability to do this is a statement of universality, but universal behaviour typically has a basin of attraction, so it should be expected that $\C$ cannot be changed completely arbitrarily. However these limitations do not apply to the required duality relation \eqref{duality-Gamma-con} since as we saw in sec. \ref{sec:duality-Gamma},
this corresponds to the special case where the form of the overall cutoff profile $C^\Lambda$ does not change, only the overall scale $\Lambda\mapsto\Lp$, and furthermore the bare interactions are perfect Wilsonian ones corresponding to a full renormalised trajectory, and thus exist at any scale.

In sec. \ref{sec:solving} we explored fully how our solutions to the reconstruction problem are related to the one-loop formula \eqref{MandR.soln} derived in ref. \cite{Manrique:2008zw}. The key was to recognise that in our second solution we employed compatible cutoffs such that when the IR cutoff meets the UV cutoff, $k\to\Lambda$, the propagator is forced to vanish identically. In sec. \ref{sec:compatible} we set out a recipe for constructing such cutoff combinations.

Although we phrased all relations in terms of a single scalar field, it is a straightforward generalisation to write the relations for multiple fields including fields with indices and those with fermionic statistics. It is therefore straightforward to generalise these relations to the case of full quantum gravity for instance. At various stages we discarded additive constant terms, but these would become background dependent. Their functional form can be determined however, and thus this would be a useful extension of this work. However, it should be borne in mind that the physical Green's functions are in any case determined by the quantum fields. 

Finally, since $S^{\Lambda}$ are perfect bare interactions, or equivalently since they are made via a tree-diagram expansion using the vertices of $\Gamma_{k=\Lambda}{}$, we can expect them to be as complicated as $\hG{k}{}$, arguably more so. 
For any large but finite $\Lambda$, we can however use $S^\Lambda$ as the starting point for constructing equally valid alternative bare actions based on either of our solutions of the reconstruction problem. 
We have already seen a small example of this in that using $S^\Lambda$ together with the standard coupling between source and fields as in \eqref{low.int.simplified} only yields a perfect action lying on a renormalised trajectory in the limit of infinite $\Lambda$, unless we impose restrictions on the source (\cf sec. \ref{sec:SandGamma}).
In fact we have an infinite dimensional space of possible bare actions to choose from (a reflection of universality). In general we can choose $\hat{\mathcal{S}}^\Lambda$ to be any action close to any point on the (infinite dimensional) critical surface containing the UV (asymptotically safe) fixed point $\hS*$, such that after appropriate tuning back into the critical surface in the limit $\Lambda\to\infty$, we again construct the renormalised trajectory (see \eg  \cite{Morris:1998}). In practice for example we can choose $\hbS\Lambda= \hS\Lambda+\sum_{i\notin{\cal R}} \alpha_i(\Lambda) {\cal O}_i$, where the sum is over the integrated irrelevant operators and $\alpha_i(\Lambda)$ are arbitrary functions of $\Lambda$ providing they remain small enough for the linearised approximation to be valid as $\Lambda\to\infty$. 


\section*{Acknowledgments} TRM acknowledges support through STFC consolidated grant no. ST/L000296/1. ZS acknowledges support through an STFC studentship.

\newpage 

\appendix

\section{Why a UV regulated effective average action must depend on the UV regulator}
\label{app:UV}

It is clear that at least for a general form of UV cutoff,  the effective average action $\hat{\Gamma}_{k}^\Lambda[\varphi]$ must depend on the UV regulator $\Lambda$ as indicated. Indeed if we embed the UV cutoff in the free propagator as done in \eqref{Gamma.flow} then the Feynman diagrams that follow from its perturbative expansion will evidently have all free propagators $1/p^2$ replaced by $\Delta^\Lambda_k(p)$. The fact that $\hat{\Gamma}_{k}^\Lambda[\varphi]$ thus depends on two scales, means that a bare action cannot be reconstructed which would directly give the continuum version $\hat{\Gamma}_k$ in the usual way. This is the first ``severe issue''  outlined above \eqref{MandR.soln}. 

Following ref. \cite{Manrique:2008zw}, a sharp UV cutoff and infrared optimised cutoff would appear to provide an exception however. With  a sharp UV cutoff in place, \eqref{R-flow} can alternatively be written
\begin{equation}
\label{R-flow-2}
	\frac{\partial}{\partial k}\hat{\Gamma}_{k}^\Lambda[\varphi]=\frac{1}											{2}\text{tr}\bigg[\bigg(R_{k}+\frac{\delta^{2}\hat{\Gamma}_{k}^\Lambda}										{\delta\varphi\delta\varphi}\bigg)^{\!-1}\frac{\partial R_{k}}{\partial k}\bigg]-\frac{1}											{2}\text{tr}\bigg[\theta(|p|-\Lambda)\bigg(R_{k}+\frac{\delta^{2}\hat{\Gamma}_{k}^\Lambda}										{\delta\varphi\delta\varphi}\bigg)^{\!-1}\frac{\partial R_{k}}{\partial k}\bigg]\,,
	\end{equation}
where the first space-time trace leads to an unrestricted momentum integral
\be 
\int \!\!\frac{d^{d}p}{(2\pi)^{d}}\, \bigg(R_{k}+\frac{\delta^{2}\hat{\Gamma}_{k}^\Lambda}										{\delta\varphi\delta\varphi}\bigg)^{\!-1}\!\!\!\!\!\!(p,-p)\,\,\frac{\partial R_{k}(p)}{\partial k}\,,
\ee
and we mean that the second  term, the ``remainder term'', has the momentum integral defining the trace restricted to $|p|>\Lambda$ as indicated. With the optimised IR cutoff profile we have $\partial R_k(p)/\partial k = 2k\theta(k^2-p^2)$ and thus, since $k\le\Lambda$, the remainder term vanishes in this case. At first sight this would appear then to allow us to consistently set $\hG{k}\Lambda[\ph] = \hG{k}{}[\ph]$ in \eqref{R-flow-2} (providing only that we restrict flows to $k\le\Lambda$), meaning that for these choice of cutoffs, the dependence of the effective average action on $\Lambda$ disappears. This is not correct however as can be seen by expanding the inverse kernel. Define the full inverse propagator as
\be 
\hat{\Delta}^{\!-1}(p) := R_k(p) +\frac{\delta^{2}\hat{\Gamma}_{k}^\Lambda}										{\delta\varphi(p)\delta\varphi(-p)}\bigg|_{\varphi=0}\,,
\ee
(temporarily suppressing the $k$ and $\Lambda$ dependence) and similarly define $\Gamma^{\prime}[\varphi]$ to be the remainder after the term quadratic in the fields is removed (which thus starts at ${\cal O}(\varphi^3)$ in a field expansion).
Then 
\bea 
\label{inverse-expanded}
\left(R_{k}+\frac{\delta^{2}\hat{\Gamma}_{k}^\Lambda}	{\delta\varphi\delta\varphi}\right)^{\!-1}\!\!\!\!\!\!(p,-p) &=& \left(\hat{\Delta}^{\!-1}+\frac{\delta^{2}\Gamma'}	{\delta\varphi\delta\varphi}\right)^{\!-1}\!\!\!\!\!\!(p,-p)\\
&=& \hat{\Delta}(p)- \hat{\Delta}(p)\frac{\delta^{2}\Gamma'	}		{\delta\varphi(p)\delta\varphi(-p)}\hat{\Delta}(p)\nonumber\\
&+&\int^\Lambda \!\!\!\!\frac{d^{d}q}{(2\pi)^{d}}\, \hat{\Delta}(p)\frac{\delta^{2}\Gamma'}{\delta\varphi(p)\delta\varphi(-p-q)}\hat{\Delta}(p+q)\frac{\delta^{2}\Gamma'}{\delta\varphi(p+q)\delta\varphi(-p)}\hat{\Delta}(p)-\cdots.\nonumber
\eea
The momentum $q$ is the external momentum injected by the fields remaining in $\Gamma'$:
\be 
\frac{\delta^{2}\Gamma'}{\delta\varphi(p)\delta\varphi(-p-q)} = \Gamma^{(3)}(p,-p-q,q;k,\Lambda)\varphi(-q)+{\cal O}(\varphi^2)\,,
\ee
where we have displayed as a simple example the 1PI three-point vertex defined as in \eqref{Gamma-vertices}. (The higher point vertices will have an integral over the field momenta with a delta-function restricting the sum to $-q$.) With a sharp UV cutoff in place, not only are the external momenta $|q|\le\Lambda$ restricted, but the momentum running through any internal line is also restricted, thus here we also have $|p+q|\le\Lambda$. This is because ultimately all the free propagators come (via Wick's theorem) from a Gaussian integral over the fields $\phi(r)$ in the path integral whose momenta $|r|\le\Lambda$ have been restricted by the sharp UV cutoff. Although the momentum $p$ already has a sharp UV cutoff $k$ provided by $\partial R_k(p)/\partial k$ which means the overall UV cutoff $\Lambda$ is invisible for it, this invisibility does not work for the other internal momenta, such as $p+q$, hidden in the construction of the inverse kernel. In other words even if the argument $p$ above is freed from its UV cutoff at $\Lambda$, this cutoff remains inside the construction in all the internal propagators, such as displayed in \eqref{inverse-expanded}, and thus despite appearances the first term on the right hand side of \eqref{R-flow-2} actually still does depend non-trivially on $\Lambda$, implying also that $\hat{\Gamma}_{k}^\Lambda[\varphi]$ is a non-trivial function of $\Lambda$.

\newpage

\bibliographystyle{hunsrt}
\bibliography{references}

\end{document}